\documentclass[twocolumn,epjc3]{svjour3}
\usepackage{graphics}
\usepackage{multicol}
\usepackage{cuted}
\usepackage{flushend}

\usepackage{microtype}
\usepackage{bbm}
\usepackage{amsmath}
\usepackage{mathtools}
\usepackage{caption}
\usepackage{subcaption}
\usepackage{amssymb}
\usepackage{mathrsfs}       
\usepackage[pdftex]{graphicx}
\usepackage{pgfplots}
\pgfplotsset{width=10cm,compat=1.9}
\usepackage{xargs}       
\usepackage{wasysym}
\usepackage{engrec}
\usepackage{enumerate}
\usepackage{appendix}
\usepackage{empheq}
\usepackage{float}
\usepackage{stmaryrd}
\usepackage[parfill]{parskip}
\usepackage[pdftex,breaklinks]{hyperref}
\usepackage{titletoc}
\usepackage{makecell}
\usepackage{lscape}
\usepackage{algorithm}
\usepackage{algorithmicx}
\usepackage{tensor}
\usepackage{algpseudocode}
\usepackage{blkarray}
\usepackage{multirow}
\usepackage{bigstrut}
\usetikzlibrary{plotmarks}
\usepackage{comment}

\newcommand{\inv}[1]{\frac 1{#1}}
\newcommand{\ket}[1]{|#1\rangle}

\newcommand{\braket}[2]{\langle#1|#2\rangle}

\newenvironment{customlegend}[1][]{%
    \begingroup
    \csname pgfplots@init@cleared@structures\endcsname
    \pgfplotsset{#1}%
}{%
    \csname pgfplots@createlegend\endcsname
    \endgroup
}%
\def\addlegendimage{\csname pgfplots@addlegendimage\endcsname}

\begin{document}
\title{Multi-reference many-body perturbation theory for nuclei}
\subtitle{II. Ab initio study of neon isotopes via PGCM and IM-NCSM calculations}

\author{M.~Frosini\thanksref{ad:saclay,em:mf} \and T.~Duguet\thanksref{ad:saclay,ad:kul,em:td}  \and J.-P.~Ebran\thanksref{ad:dam,ad:fakedam,em:jpe} \and B.~Bally\thanksref{ad:dft,em:bb} 
\and T.~Mongelli\thanksref{ad:tud,em:tm}
\and T.R.~Rodr\'iguez\thanksref{ad:dft,ad:cdiaf,em:tr} \and R.~Roth\thanksref{ad:tud,ad:hfhf,em:rr}  \and V.~Som\`a\thanksref{ad:saclay,em:vs}}

\thankstext{em:mf}{\email{mikael.frosini@cea.fr}}
\thankstext{em:td}{\email{thomas.duguet@cea.fr}}
\thankstext{em:jpe}{\email{jean-paul.ebran@cea.fr}}
\thankstext{em:bb}{\email{benjamin.bally@uam.es}}
\thankstext{em:tm}{\email{tobias.mongelli@physik.tu-darmstadt.de}}
\thankstext{em:tr}{\email{tomas.rodriguez@uam.es}}
\thankstext{em:rr}{\email{robert.roth@physik.tu-darmstadt.de}}
\thankstext{em:vs}{\email{vittorio.soma@cea.fr}}

\institute{
\label{ad:saclay}
IRFU, CEA, Universit\'e Paris-Saclay, 91191 Gif-sur-Yvette, France 
\and
\label{ad:kul}
KU Leuven, Department of Physics and Astronomy, Instituut voor Kern- en Stralingsfysica, 3001 Leuven, Belgium 
\and
\label{ad:dam}
CEA, DAM, DIF, 91297 Arpajon, France
\and
\label{ad:fakedam}
Universit\'e Paris-Saclay, CEA, Laboratoire Mati\`ere en Conditions Extr\^emes, 91680 Bruy\`eres-le-Ch\^atel, France
\and
\label{ad:dft}
Departamento de F\'isica Te\'orica, Universidad Aut\'onoma de Madrid, 28049 Madrid, Spain 
\and
\label{ad:cdiaf}
Centro de Investigaci\'on Avanzada en F\'isica Fundamental-CIAFF-UAM, 28049 Madrid, Spain 
\and
\label{ad:tud}
Institut f\"ur Kernphysik, Technische Universit\"at Darmstadt, 64289 Darmstadt, Germany
\and
\label{ad:hfhf}
Helmholtz Forschungsakademie Hessen f\"ur FAIR, GSI Helmholtzzentrum, 64289 Darmstadt, Germany
}

\date{Received: \today{} / Revised version: date}

\maketitle
%
%
\begin{abstract}
The neon isotopic chain displays a rich phenomenology, ranging from clustering in the ground-state of the self-conjugate doubly open-shell stable $^{20}$Ne isotope to the physics of the island of inversion around the neutron-rich $^{30}$Ne isotope. This second (i.e. Paper II) of the present series proposes an extensive ab initio study of neon isotopes based on two complementary many-body methods, i.e. the quasi-exact in-medium no-core shell model (IM-NCSM) and the projected generator coordinate method (PGCM) that is ideally suited to capturing strong static correlations associated with shape deformation and fluctuations. Calculations employ a state-of-the-art generation of chiral effective field theory Hamiltonians and evaluate the associated systematic uncertainties.  In spite of missing so-called dynamical correlations, which can be added via the multi-reference perturbation theory proposed in the first paper (i.e. Paper I) of the present series~\cite{frosini21b}, the PGCM is shown to be a suitable method to tackle the low-lying spectroscopy of complex nuclei. Still, describing the physics of the island of inversion constitutes a challenge that seems to require the inclusion of dynamical correlations. This is addressed in the third paper (i.e. Paper III) of the present series~\cite{frosini21d}.
\end{abstract}

\section{Introduction}
\label{intro}

The projected generator coordinate method (PGCM) based on the mixing of Bogoliubov vacua generated by solving constrained Hartree-Fock-Bogoliubov (HFB) mean-field equations has been traditionally employed with empirical effective interactions~\cite{bender03b,Niksic:2011sg,Robledo:2018cdj}. In spite of being ill-defined and affected by potentially dangerous spurious contaminations~\cite{Dobaczewski:2007ch,Duguet:2008rr,Bender:2008rn,Lacroix:2008rj} in this particular context, such PGCM calculations have been successfully applied to describe numerous nuclear phenomena over the last few decades.

The PGCM has also been employed in the context of so-called valence-space calculations based on appropriate effective interactions~\cite{Gao:2015dla,Jiao:2017opc,Shimizu:2021ltl}, although less often. Employing sophisticated realizations of the PGCM ansatz, solutions obtained from an exact diagonalization for pf-shell Ca isotopes~\cite{Bally:2019miu} or the complete set of sd-shell nuclei~\cite{Sanchez-Fernandez:2021nfg} have recently been shown to be accurately reproduced. These works demonstrate the capacity of the PGCM to efficiently capture strong {\it static} correlations emerging within a small energy window around the Fermi energy. 

In the present work, the PGCM is utilized in the context of {\it ab initio} calculations  that aim at approximating exact solutions of Schr{\"o}dinger's equation in the complete  $A$-body Hilbert space based on realistic nuclear Hamiltonians rooted into quantum chromodynamics. PGCM calculations have already been performed recently on the basis of realistic Hamiltonians that were pre-processed via unitary similarity renormalization group (SRG) transformations~\cite{Frosini:2021tuj} and possibly further pre-processed via unitary in-medium SRG (IMSRG) transformations~\cite{Yao:2019rck,Yao:2018qjv}. However, and independently of the pre-processing of the Hamiltonian, the PGCM is not amenable to an exact solution of $A$-body Schr{\"o}dinger's equation. Indeed, while very efficient at grasping strong static (collective) correlations, the PGCM is not suited to capture weak, so-called dynamical, correlations\footnote{In valence space calculations, dynamical correlations are essentially accounted for, at least in principle, through the effective Hamiltonian. This is the reason why PGCM can well reproduce exact solutions in this particular context. See Sec.~\ref{spectroNechain} for an illustration of this feature.}. The key novelty of the present work relates to the formulation of a multi-reference perturbation theory (PGCM-PT) formalism~\cite{frosini21b} embedding, for the first time,  the PGCM into a genuine ab initio expansion method capable of grasping dynamical correlations in a systematic fashion. 

While the PGCM-PT formalism was explained in detail in Ref.~\cite{frosini21b}, hereafter coined as Paper I, the present paper (Paper II) is devoted to presenting numerical results obtained through its leading order, i.e. PGCM, reduction. Specifically focusing on even-even neon isotopes, the objectives of this work are to
\begin{enumerate}
\item benchmark PGCM calculations against quasi-exact results obtained via the so-called in-medium no core shell model (IM-NCSM),
\item deliver ab initio predictions of spectroscopic properties of even-even Ne isotopes,
\item gauge uncertainties and convergence of the many-body results associated with the order-by-order chiral effective field theory ($\chi$EFT) expansion of the Hamiltonian.
\end{enumerate}
Based on the above results, the following paper~\cite{frosini21d}, i.e. Paper III, will present the first PGCM-PT calculations beyond zeroth order and characterize the way absolute and relative PGCM energies are amended by the inclusion of dynamical correlations.

The present paper is organized as follows. All the ingredients of the calculations (Hamiltonians, many-body formalisms, numerical settings, uncertainty evaluations) are detailed in Sec.~\ref{sec:1} whereas a large body of results is presented in Sec.~\ref{sec_results}. The conclusions of the present work are then given in Sec.~\ref{sec_conclusions}. Eventually, four technical appendices complement the body of the paper.

\section{Many-body calculations}
\label{sec:1}

\subsection{Nuclear Hamiltonian}

The present calculations employ the family of $\chi$EFT Hamiltonians $H$ introduced in Ref.~\cite{Huther_2020} and constructed at next-to-leading (NLO), next-to-next-to-leading (N$^2$LO) and next-to-next-to-next-to-leading (N$^3$LO) orders according to Weinberg's power counting~\cite{Epelbaum:2008ga,Epelbaum:2019jbv,Machleidt:2020vzm}. The same non-local regulators and cut-off values ($\Lambda = 500$\,MeV) are employed in the two-nucleon and three-nucleon sectors; see Refs.~\cite{Huther_2020,Entem:2017gor} for the details of the fitting protocol. This family of interactions was shown to robustly reproduce selected experimental energies and radii from p-shell nuclei to nickel isotopes and to resolve several deficiencies of the previous generations of $\chi$EFT Hamiltonians.

To be employed in the many-body calculations, the $\chi$EFT Hamiltonians are evolved to a lower resolution scale \(\lambda_{\text{srg}}\) via vacuum SRG transformations~\cite{Bogner:2009bt,PhysRevLett.107.072501,PhysRevC.90.024325} while discarding induced operators beyond three-body terms. The values of \(\lambda_{\text{srg}}\) employed in the  many-body calculations presented below are specified later on.

\subsection{PGCM}

The PGCM presently employed has been described in Paper I~\cite{frosini21b} and the reader is referred to it for details. 

\subsubsection{Choice of collective coordinates}

The PGCM state relies on a set $\text{B}_{q} \equiv \{ \ket{\Phi(q)}; q \in \text{set} \}$ of Bogoliubov states differing by the value of the (typically multi-dimensional) collective deformation parameter $q$ and obtained by repeatedly solving constrained Hartree-Fock-Bogoliubov equations. As an intermediate step, the calculation thus delivers a HFB total energy surface (TES) as a function of $q$.

Typically, $q$ presently collects quadrupole $(q_{2\mu})$ and axial octupole $(q_{30})$ moments, i.e.
\begin{subequations}
\label{moments}
\begin{align}
Q_{\lambda \mu} &\equiv r^{\lambda} Y^{\lambda}_{\mu}(\theta,\varphi) \, , \\
q_{\lambda \mu} & \equiv \frac12 \langle \Phi(q) | Q_{\lambda \mu} + (-1)^\mu  Q_{\lambda -\mu}  | \Phi(q)  \rangle \, ,  
\end{align}
\end{subequations}
where $Y^{\lambda}_{\mu}(\theta,\varphi)$ is a spherical harmonic of degree $\lambda$ and
order $\mu$, such that
\begin{subequations}
\label{moments2}
\begin{align}
q_{20} &\equiv \langle \Phi(q) | Q_{20} | \Phi(q)  \rangle \, , \\
q_{21} &\equiv \frac{1}{2} \langle \Phi(q) | Q_{21} - Q_{2-1} | \Phi(q)  \rangle \, , \\
q_{22} &\equiv \frac{1}{2} \langle \Phi(q) | Q_{22} + Q_{2-2} | \Phi(q)  \rangle \, , \\
q_{30} &\equiv \langle \Phi(q) | Q_{30} | \Phi(q)  \rangle \, .
\end{align}
\end{subequations}

In the present calculations, $q_{10}$ and $q_{11}$ are set to zero to avoid the spurious motion of the nucleus center of mass. Similarly, $q_{21}$ is set to zero to fix the orientation of the nucleus. From the moments, one introduces deformation parameters according to
\begin{subequations}
\label{deformations}
\begin{align}
\beta_{2} &\equiv \frac{4\pi}{(3 R^{2} A)} \sqrt{q^2_{20}+2q^2_{22}} \, , \\
\gamma_{2} &\equiv \arctan\left(\frac{\sqrt{2}q_{22}}{q_{20}}\right) \, , \\
\beta_{3} &\equiv \frac{4\pi}{(3 R^{3} A)} q_{30} \, , 
\end{align}
\end{subequations}
with $R\equiv 1.2 A^{1/3}$ and $A\equiv N+Z$ the mass number. Whenever the deformation is purely axial, $\beta_{2}$ reduces to the traditional axial quadrupole deformation parameter.

Each Bogoliubov state $\ket{\Phi(q)}$ is further projected, whenever necessary, onto good symmetry quantum numbers \(\sigma\equiv(\text{J} \text{M} \Pi \text{N} \text{Z})\equiv (\tilde{\sigma}M)\) \cite{Bally21a}, i.e. onto total angular momentum $J$ and projection $M$, parity $\Pi=\pm 1$ as well as neutron $N$ and proton $Z$ numbers. This procedure generates a set $\text{PB}_{q\tilde{\sigma}}$ of {\it projected} Bogoliubov states for each realization $\tilde{\sigma}$ of the symmetry quantum numbers and an associated projected HFB (PHFB) TES.

Eventually, the PGCM ansatz mixes all the states belonging to $\text{PB}_{q\tilde{\sigma}}$. The unknown coefficients $\{ f^{\tilde{\sigma}}_{\mu}(q); q \in \text{set} \}$ of the linear combination are determined via the application of Ritz' variational principle. This leads to solving Hill-Wheeler-Griffin's (HWG) equation\footnote{The diagonalization is performed separately for each value of $\tilde{\sigma}$.} \cite{Hill53a,Griffin57a} that is nothing but a generalized eigenvalue problem represented in the set $\text{PB}_{q\tilde{\sigma}}$ of non-orthogonal PHFB states. The practical aspect of dealing with the linear redundancies associated with the non-orthogonality of the PHFB states when solving HWG's equation are briefly discussed in App.~\ref{linear-redund_HWG}.

\subsubsection{Numerical setting}

In the present paper, two sets of HFB~\cite{Bally:2021kfw,frosini21e} and HWG~\cite{bally21b,frosini21f} solvers are used. While the first set~\cite{Bally:2021kfw,bally21b} offers more flexibility regarding the enforced/relaxed symmetries in the computation of the HFB states and operator kernels entering HWG's equation, the second set~\cite{frosini21e,frosini21f} can exactly handle three-nucleon interactions. 

Based on these solvers, the calculations performed in the present study involve
\begin{enumerate}
\item the potential breaking of
\begin{enumerate}
\item global neutron and proton gauge symmetries,
\item rotational symmetry,
\item parity,
\end{enumerate}
\item the associated restoration of
\begin{enumerate}
\item $N$ and $Z$,
\item $J$ and $M$,
\item $\Pi$,
\end{enumerate}
\item constraints associated with 
\begin{enumerate}
\item axial quadrupole ($q_{20}$),
\item non-axial quadrupole  ($q_{2\pm2}$),
\item axial octupole ($q_{30}$),
\end{enumerate}
operators.
\end{enumerate}

Calculations are performed using a spherical harmonic oscillator (HO) basis of the one-body Hilbert space ${\cal H}_1$. The basis is characterized by the value of the oscillator frequency $\hbar\omega$ and by the number of oscillator shells kept in the calculations. The latter is parameterized by the quantity $e_{\text{max}} = \text{max}(2n+l)$, where $n$ and $l$ respectively denote the principal quantum number and the orbital angular momentum of the basis states.

When representing $n$-body operators, the natural truncation of the tensor-product basis of the $n$-body Hilbert space ${\cal H}_n$ is set by $e_{n\text{max}}\equiv ne_{\text{max}}$. One and two-body operators are thus represented using $e_{1\text{max}}=e_{\text{max}}$ and $e_{2\text{max}}=2e_{\text{max}}$, respectively. However, the fixed value $e_{3\text{max}}=14$ ($\ll3e_{\text{max}}$) is used to represent the three-nucleon interaction given that employing $3e_{\text{max}}$ for workable values of $e_{\text{max}}$ is largely beyond today's capacities\footnote{A novel framework capable of handling values up to $e_{3\text{max}}=28$ in the normal-ordered two-body approximation was proposed very recently~\cite{Miyagi21a}. However, as discussed later, the present truncation of $e_{3\text{max}}=14$ is sufficient to produce converged results for the light nuclei under present consideration .}.

\subsubsection{Uncertainties}
\label{uncertainPGCM}

The uncertainties of PGCM calculations are of several origins and nature\footnote{When adding two uncertainties $\sigma_1$ and $\sigma_2$, presently supposed to be uncorrelated, the total one is computed as  $\sigma_{\text{tot}} \equiv \sqrt{\sigma^2_1+\sigma^2_2}$. When considering an observable $O = O_1 - O_2$, e.g. an excitation energy, its uncertainty is computed under the hypothesis that the uncertainties associated with $O_1$ and $O_2$ are fully correlated, i.e. using  $\sigma_{O} \equiv |\sigma_{O_1}-\sigma_{O_2}|$.}
\begin{itemize}
\item Numerical representation
\begin{itemize}
\item {\it Model-space truncation}. Results depend on the choice of the one-body basis parameters $(\hbar\omega,e_{\text{max}})$ and the truncation of three-body operators $e_{3\text{max}}$. While the nominal results discussed in the following are obtained for $(\hbar\omega,e_{\text{max}},e_{3\text{max}})=(12,10,14)$, the associated uncertainty is evaluated in each nucleus according to a procedure described in Sec.~\ref{basis_optimization} and typically included in the error bars displayed in several of the figures below.
\item {\it Approximate three-body interaction}. While three-body interaction terms can be handled exactly, doing so typically increases the runtime of PGCM calculations by three orders of magnitude compared to using a two-nucleon interaction only~\cite{frosini21e,frosini21f}. In order to avoid this significant cost, a novel operator rank-reduction method that generalizes the so-called normal-ordered two-body (NO2B) approximation was recently introduced~\cite{Frosini:2021tuj}. The approximation was shown to induce errors below $2-3\%$ across a large range of nuclei, observables and many-body methods when employing low-resolution Hamiltonians as done in the present work. Specifically, the PGCM errors on the ground-state charge radius and the low-lying excitation energies of $^{20}$Ne ($^{30}$Ne) were shown to be of $0.7\%$ ($2.5\%$) and $1.5\%$ ($2.6\%$), respectively. While not included in the error bars appearing in some of the figures below, a conservative $2-3\%$ error is to be appropriately attributed.
\item {\it Discretization errors}. PGCM results depend on the discretization of the employed generator coordinate(s) and on the procedure described in App.~\ref{linear-redund_HWG} to handle linear redundancies when solving HWG equation. The dependence of our PGCM results on these two numerical parameters have been checked and found to be negligible compared to the other sources of uncertainty. 
\end{itemize}
\item Many-body expansion
\begin{itemize}
\item {\it Generator coordinates}. PGCM results depend on the choice of generator coordinates employed in the calculation. While it is hard to envision a systematic way to evaluate an associated uncertainty, the dependence of the results on the  generator coordinates that are expected to be dominant is gauged by generating results (a) with or without the octupole degree of freedom and (b) with or without the triaxial degree of freedom.
\item {\it Many-body truncation}. Given a PGCM ansatz, the PGCM-PT formalism developed in Paper I allows one to embed it into a systematic many-body expansion that, at least in principle, converge towards the exact solution. Constituting the leading order contribution to the expansion, PGCM results carry an uncertainty associated with the corresponding truncation. Because it is the goal of Paper III to present the first computation of the next correction, i.e. PGCM-PT(2), the associated uncertainty is not evaluated in the present paper but simply commented on at various points below. 
\end{itemize}
\item Hamiltonian
\begin{itemize}
\item {\it $\chi$EFT truncation}. The hierarchy of terms in the chiral expansion allows us to increase the precision at each order and coherently assess truncation errors. These errors are consistently propagated to many-body calculations and are to be added to the errors coming from the many-body method itself. The uncertainty of a many-body observable \(X\) at N$^2$LO and N$^3$LO reads~\cite{Huther_2020,LENPIC:2018lzt,Epelbaum:2014efa} 
 \begin{subequations}
    \begin{align}
    \delta X_{N^2LO} &\equiv 
    Q |X_{N^2LO} - X_{NLO}|\,,  \\
    \delta X_{N^3LO} &\equiv \max\left[
    Q |X_{N^3LO} - X_{N^2LO}|, \right. \nonumber 
    \\ &\phantom{\equiv \max[} \left. Q^2 |X_{N^2LO} - X_{NLO}|
    \right], 
    \end{align}
 \end{subequations}
where the expansion parameter \(Q\) denotes the ratio of a typical momentum scale characterizing medium-mass nuclei over the $\chi$EFT breakdown scale. The value \(Q=1/3\) is presently employed; see Ref.~\cite{Huther_2020,LENPIC:2018lzt,Epelbaum:2014efa} for details.
\item  {\it SRG dependence}. The vacuum SRG transformation induces an intrinsic error associated with the violation of unitarity due to neglected induced operators beyond three-body terms. Furthermore, the uncertainty associated with the truncation of the many-body expansion itself depends on the transformation, which is typically minimized by working with low-resolution Hamiltonians as done in the present work. Overall, this induces a dependence of the results on the SRG parameter $\lambda_{\text{srg}}$. While the nominal results are systematically provided for $\lambda_{\text{srg}}=1.88$\,fm$^{-1}$, the variation of the results obtained for $\lambda_{\text{srg}}=2.23$\,fm$^{-1}$ will be quoted to provide an idea of the sensitivity of selected observables. 
\end{itemize}
\end{itemize}
Eventually, only model-space (inner error bars) and interaction (total error bars) uncertainties are reported in the figures containing PGCM calculations.

\subsection{IM-NCSM}

As a complement and a benchmark of the PGCM calculations, the IM-NCSM approach~\cite{Gebrerufael:2016xih} is used to describe even-even neon isotopes. 

\subsubsection{Methodology}

The IM-NCSM starts by pre-processing (already SRG-evolved) operators $O$ through a nucleus-dependent unitary MR-IMSRG transformation $U(s)$ parameterized\footnote{Following standard conventions, the vacuum SRG is characterized by the variable $\lambda$ in fm$^{-1}$ whereas the in-medium SRG is parameterized by the variable $s$ in MeV$^{-1}$. While $\lambda$ decreases from infinity towards zero throughout the evolution, the variable $s$ does the opposite.} by the real variable $s$. The lowest eigenstate with appropriate symmetry quantum numbers obtained from a prior NCSM calculation in a small reference space including all basis Slater determinants with up to $N^{\text{ref}}_{\text{max}}$ HO excitation quanta above the lowest-energy basis states serves as a multi-configurational reference state. The transformed operator $O(s)$ expressed in normal-ordered form~\cite{kut97a,kong10a} with respect to the NCSM reference state is truncated  beyond two-body operators, i.e. at the MR-IMSRG(2) level, which induces a breaking of unitarity that needs to be monitored.

The transformation $U(s)$ is tailored to suppress the terms of the pre-processed Hamiltonian $H(s)$ that couple the NCSM reference space to the rest of the Hilbert space. This decoupling corresponds to the incorporation of dynamical correlations into the transformed Hamiltonian. This leads to an extremely fast convergence of a subsequent NCSM calculation as a function of the truncation parameter $N_{\text{max}}$ for appropriate values of the flow parameter $s$. This final NCSM calculation, performed with $H(s)$, directly yields the ground and excited-state energies as eigenvalues. Using the NCSM eigenvectors and the consistently evolved operators $O(s)$ the other relevant observables are computed, including non-scalar quantities like magnetic dipole or electric quadrupole moments and transition strengths \cite{Vobig:2021,Mongelli:2021}.

\subsubsection{Numerical setting}

All IM-NCSM calculations are performed with a natural orbital basis constructed from a perturbatively corrected density matrix for the isotope of choice \cite{Tichai:2018qge}. This leads to a fast and frequency-independent convergence of NCSM calculations, as shown in \cite{Tichai:2018qge}, and, thus, improves the reference state for the IM-NCSM and allows us to limit all calculations to a single frequency $\hbar\omega=20$\,MeV of the underlying oscillator basis. While the initial NCSM calculation is performed for $N^{\text{ref}}_{\text{max}}$ = 0 or 2, the final NCSM calculation goes up to $N_{\text{max}}=4$, being fully converged.

In the MR-IMSRG part of the calculation, $e_{\text{max}}=12$ and $e_{3\text{max}}=14$ are employed. The flow equations rely on a modified version of the so-called White generator~\cite{Hergert:2016iju,Vobig:2021,Mongelli:2021} adapted to the $N_{\text{max}}$-truncated reference space and the value of the flow-parameter is chosen large enough to warrant convergence of the evolved Hamiltonian, i.e. typically around $s = 80$.

A shifted center-of-mass HO Hamiltonian is consistently evolved through the MR-IMSRG and added to the Hamiltonian with a small pre-factor $\lambda_{\text{cm}} = 0.2$ to remove spurious center-of-mass excitations from the low-lying spectrum. 

\begin{figure}
    \centering
    \includegraphics[width=.55\textwidth]{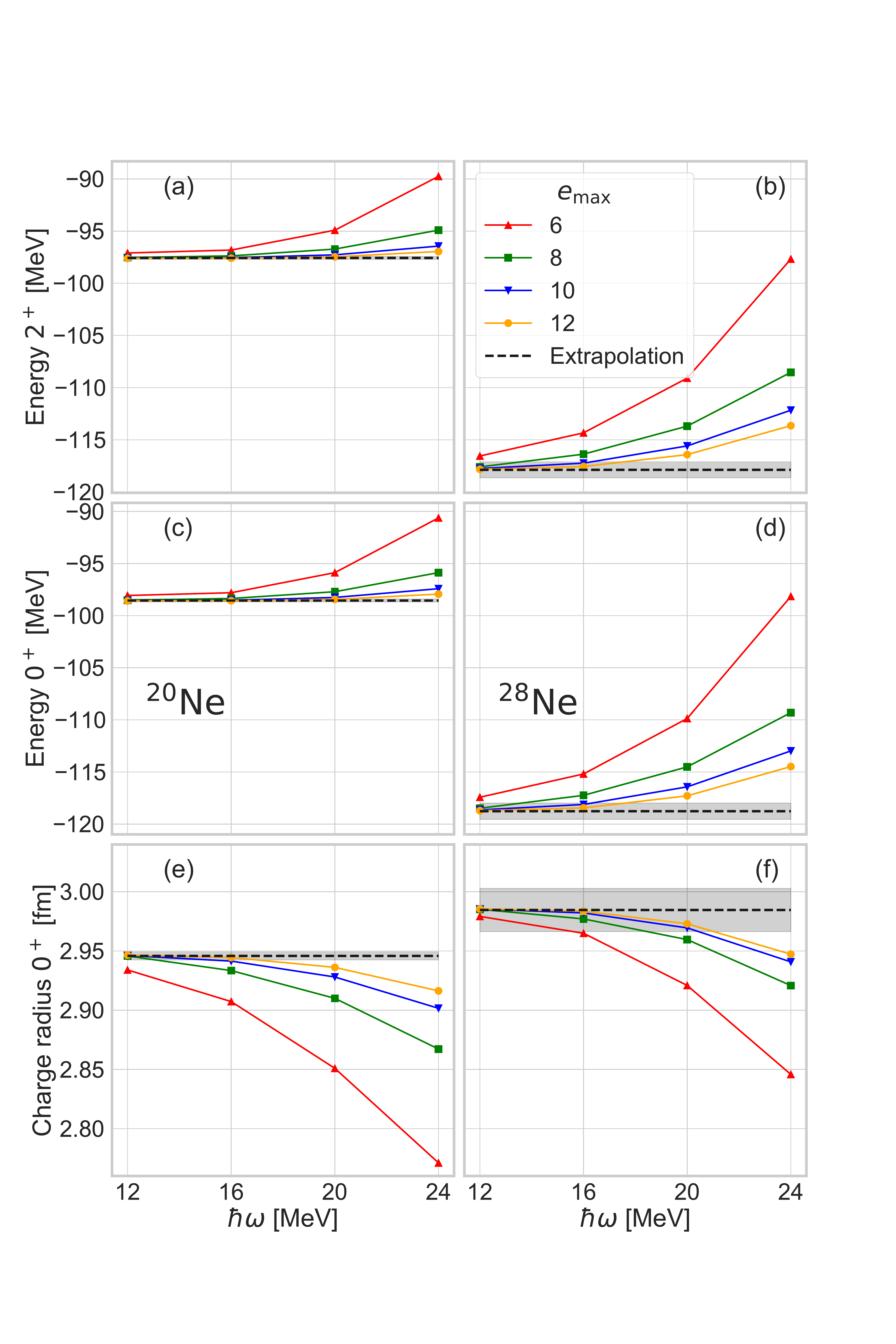}
    \caption{Dependence of PHFB results in $^{20}$Ne (left column) and $^{28}$Ne (right column) on the employed HO model space. Results are plotted as a function of $\hbar\omega$ for various values of \(e_{\text{max}}\). The dashed lines denote extrapolated values whereas the grey band provides the associated uncertainty. The first row (panels (a) and (b)) focuses on the first $2^+$ absolute energy whereas the second (panels (c) and (d)) and third (panels (e) and (f)) rows provide the ground-state energy and associated rms charge radius. Calculations employ the N$^3$LO $\chi$EFT Hamiltonian with $\lambda_{\text{srg}}=1.88 $\,fm$^{-1}$.}
    \label{fig:hwemax}
\end{figure}

\subsubsection{Uncertainties}

The nominal IM-NCSM results quoted below are obtained for the largest NCSM model spaces  $N^{\text{ref}}_{\text{max}}=2$, $N_{\text{max}}=4$ and $s \sim 80$. The uncertainties quoted for all results are constructed in the following way:
\begin{itemize}
\item  {\it Many-body}. Many-body uncertainties are estimated from explicit variations of the relevant truncation parameters, i.e., we repeat the IM-NCSM calculation using $N^{\text{ref}}_{\text{max}}=0$, $N_{\text{max}}=2$ or $s\sim 40$, varying only one parameter at a time and using the maximum deviation from the nominal calculation to estimate the uncertainty. These errors are typically dominated by the effect of $N^{\text{ref}}_{\text{max}}$ although the dependence on the flow parameter is sometimes not negligible\footnote{For $N_{\text{max}}$-converged calculations, the dependence of the end results on $s$ probes the effects of truncating the transformed Hamiltonian to the normal-ordered two-body level throughout the MR-IMSRG evolution based on the reference NCSM state.}.
\item  {\it Hamiltonian}. The chiral order-by-order uncertainties are extracted from a pointwise Bayesian model~\cite{Mel19,Maris:2020qne} and we quote 68\% degree-of-believe intervals. The resulting error bars are comparable to the ones obtained from the simpler difference scheme used for the PGCM calculations. The uncertainty associated with the initial SRG parameter $\lambda_{\text{srg}}$ is not shown explicitly, for the ground-state energies, e.g., the dependence on this flow parameter is smaller than the other uncertainties.
\end{itemize}

The IM-NCSM results reported in the following figures show the many-body uncertainties as inner error-bars and the total uncertainties, obtained as a simple sum of many-body and interaction uncertainties, as shaded bands.

\section{Results}
\label{sec_results}

\subsection{Model-space convergence}
\label{basis_optimization}

The convergence of PHFB results with respect to the employed HO model space has been checked for all Ne isotopes. In this test, the HFB minimum in the $(q_{20},q_{30}$) plane, systematically obtained at $\beta_{3}=0$ (see Sec.~\ref{TES} below), is projected on good neutron and proton numbers as well as on the desired angular momentum $J$. Results for two representative examples, $^{20}$Ne and $^{28}$Ne, are displayed in Fig.~\ref{fig:hwemax} for the ground-state energy and the root-mean-square (rms) charge radius\footnote{Charge radii are computed starting from point-proton radii and correcting for the finite charge distributions of protons and neutrons, in addition to the Darwin-Foldy term (see, e.g., Ref.~\cite{Soma21} for details).}, as well as for the absolute energy of the first $2^{+}$ state. 

The three observables show a typical convergence pattern consisting of curves that gradually become independent of $\hbar \omega$ and closer to each others as the basis size increases. At each step of the way, the HO frequency delivering the least sensitive results to  \(e_{\text{max}}\), i.e. the results that are closest to the converged value, is given by $\hbar \omega =12$\,MeV. Taking the least favorable case, i.e. $^{28}$Ne, the energy of the first $0^+$ ($2^+$) changes by $70$\,keV ($72$\,keV) when going from \(e_{\text{max}}=10\) to \(e_{\text{max}}=12\) whereas the ground-state charge radius increases by $10^{-4}$\,fm. Taking the results displayed in Fig.~\ref{fig:hwemax} for $\hbar\omega \geq 12$\,MeV, their infra-red extrapolation towards the infinite basis limit is performed according to the procedure described in Ref.~\cite{Furnstahl:2012qg} for both energies and radii. The result of the extrapolation is also displayed, along with its uncertainty, in Fig.~\ref{fig:hwemax}.
 
\begin{figure}
    \centering
    \includegraphics[width=.45\textwidth]{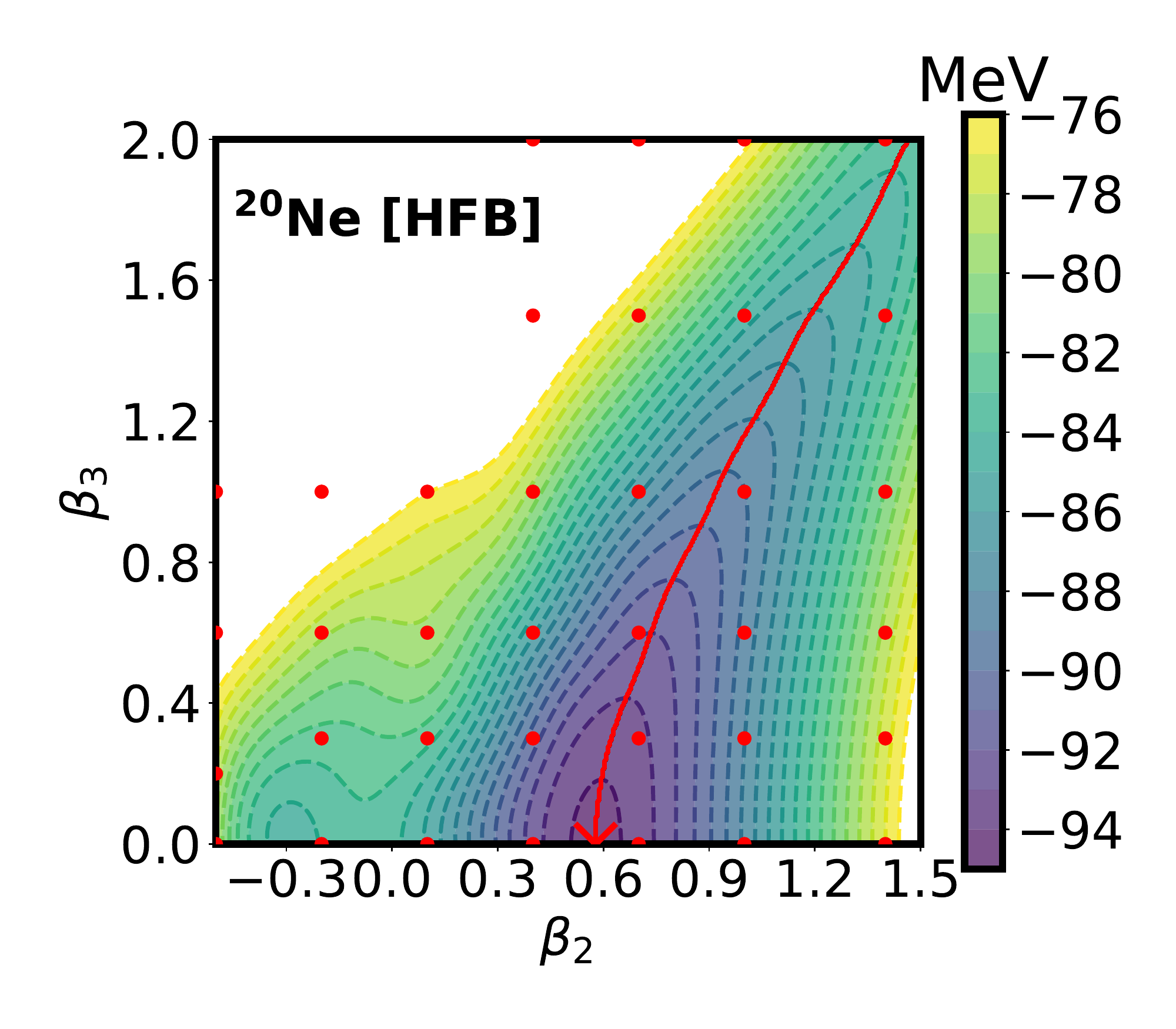}
    \caption{(Color online) Constrained HFB TES  of $^{20}$Ne in the axial $(\beta_2, \beta_3)$ plane. The (red) full line indicates the lowest-energy path, with the arrow positioned at the minimum of the TES. The (red) dots characterize the set of HFB states used in the subsequent PGCM calculation. Calculations employ the N$^3$LO $\chi$EFT Hamiltonian with $\lambda_{\text{srg}}=1.88 $\,fm$^{-1}$.}
    \label{fig:HFBTES}
\end{figure}

\begin{figure*}
    \centering
    \includegraphics[width=1.05\textwidth]{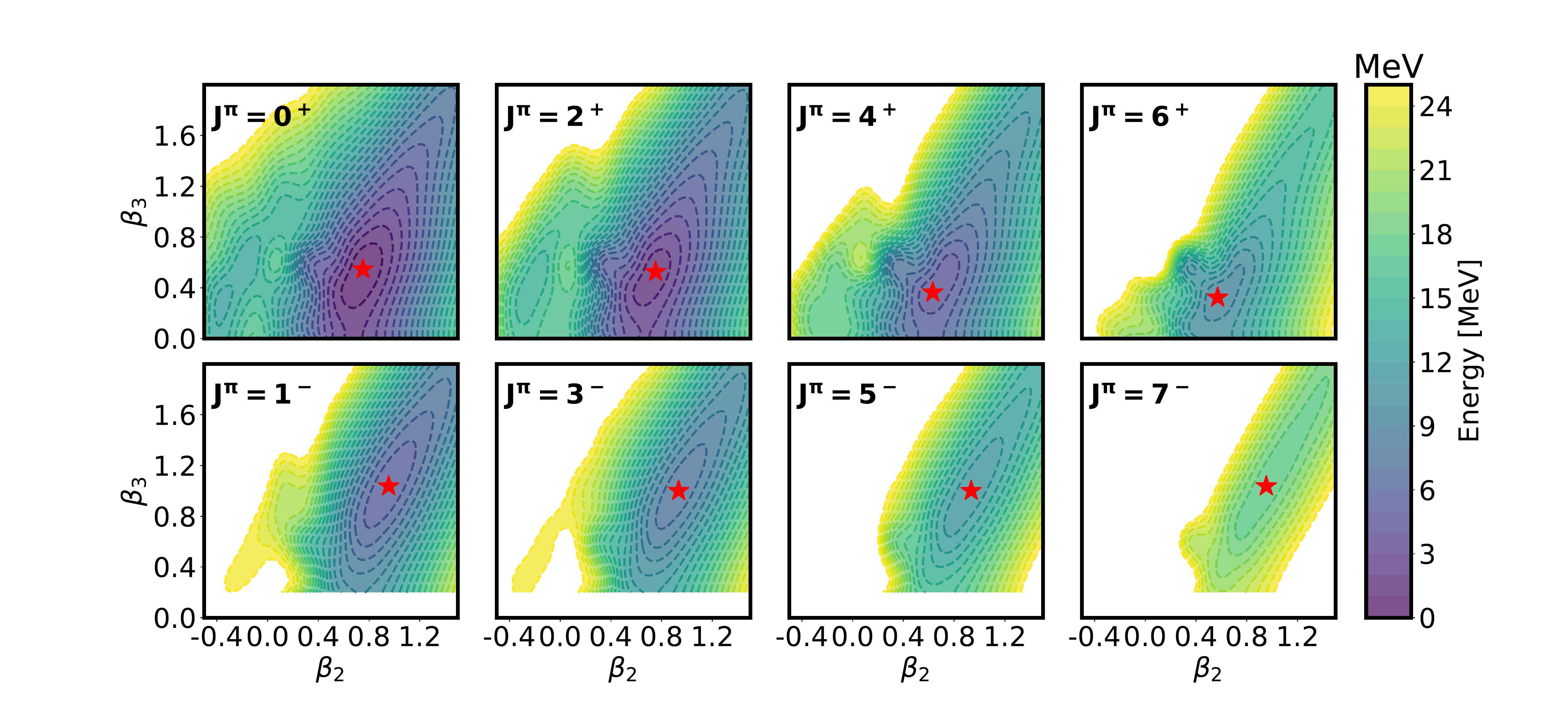}
    \caption{(Color online) Projected HFB TES  of $^{20}$Ne in the axial $(\beta_2, \beta_3)$ plane for  spin-parity values $J^{\pi}=0^{+}, 1^{-}, 2^{+}, \ldots , 7^{-}$. In each case, the minimum of the TES is indicated by a (red) star. Calculations employ the N$^3$LO $\chi$EFT Hamiltonian with $\lambda_{\text{srg}}=1.88 $\,fm$^{-1}$.}
    \label{fig:PHFBTES}
\end{figure*}

All PGCM results presented in the following have been obtained for $(\hbar\omega,e_{\text{max}},e_{3\text{max}})=(12,10,14)$. In most of the figures shown below, these nominal values are displayed with an error bar associated with the model space convergence. The error adds, in the sense explained in footnote 5, the difference between the computed result and the extrapolated value to the uncertainty on this extrapolated value. Focusing again on the least favorable case, i.e. $^{28}$Ne, model-space uncertainties on the nominal energy of the first $0^+$ and $2^+$ states are $830$\,keV ($0.7\%$) and $810$\,keV ($0.7\%$), respectively, whereas the uncertainty on the ground-state charge radius is $0.02$\,fm ($0.7\%$).

Furthermore, the impact of $e_{3\text{max}}$ has been studied by varying the truncation parameter in the range $e_{3\text{max}} = 8-14$ for selected observables. 
Overall, both energies and radii are found to be well converged with respect to $e_{3\text{max}}$, with changes between $e_{3\text{max}} = 12$ and 14 amounting in the least favorable cases to 2-300~keV for total binding energies and $10^{-3}$\,fm for charge radii. These uncertainties can be thus effectively incorporated in the larger ones resulting from the infinite-basis extrapolation discussed above.

Given that model-space uncertainties tend to cancel out in excitation spectra, the errors on the latter are typically smaller than for absolute energies. One must note that model-space uncertainties of the nominal calculations are sub-leading compared to the error associated with the rank-reduction of the three-nucleon interaction whose maximal value along the Ne chain has been evaluated to be respectively $2.5\%$ and $2.6\%$ for the ground-state charge radius and low-lying excitation energies of $^{30}$Ne~\cite{Frosini:2021tuj}. 

\subsection{$^{20}$Ne}
\label{20Ne}

The present study focuses first on the stable $^{20}$Ne isotope. This nucleus has been extensively studied experimentally and theoretically in the past~\cite{zhou16a,Marevi__2018}, in part because it is one of the few nuclei displaying a strong admixture of cluster configurations in the ground state. The ab initio description of this doubly open-shell nucleus is thus a challenge given that it is necessary to appropriately capture both dynamical and static correlations.

\subsubsection{Total energy surfaces}
\label{TES}

Figure~\ref{fig:HFBTES} displays the HFB TES of $^{20}$Ne in the axial $(\beta_2, \beta_3)$ plane. The energy minimum is found for the reflection-symmetric prolate shape characterized by deformation parameters ($\beta_2 =0.57$, $\beta_3 =0$). Still, the TES is more shallow in the octupole direction than in the quadrupole direction such that one may anticipate octupole shape fluctuations in the ground-state and an octupole vibration at an energy lower than the quadrupole one.

Figure~\ref{fig:PHFBTES} shows the PHFB TES in the axial $(\beta_2, \beta_3)$ plane for spin-parity  $J^{\pi}=0^{+}, 1^{-}, 2^{+}, \ldots, 7^{-}$. Each HFB state is projected onto neutron and proton numbers $(N, Z) =(10, 10)$ using $N_{\varphi_n}=N_{\varphi_p}=7$ mesh points in the interval $\varphi_{n,p} \in [0,\pi]$. The projection on good angular momentum involves $N_{\beta}=20$ Euler angles in the interval $\varphi_{\beta} \in [0,\pi]$. Static correlations associated with symmetry restorations favor deformed configurations in both $\beta_2$ and $\beta_3$ directions for both positive- and negative-parity states. The minimum of the $0^+$ TES is thus located at ($\beta_2 =0.75$, $\beta_3 =0.53$) and the TES is softer along both $\beta_2$ and  $\beta_3$ directions than at the HFB level. With increasing $J$, the energy minimum of positive-parity states becomes more stable but drifts to configurations with smaller multipole moments. The minimum of the $1^-$ TES is located at larger deformations ($\beta_2 =0.93$, $\beta_3 =1.0$) in the $0^+$ one. While  the minimum also becomes more stable with increasing $J$, it however remains at the same deformations.

\begin{figure*}
    \centering
    \includegraphics[width=0.85\textwidth]{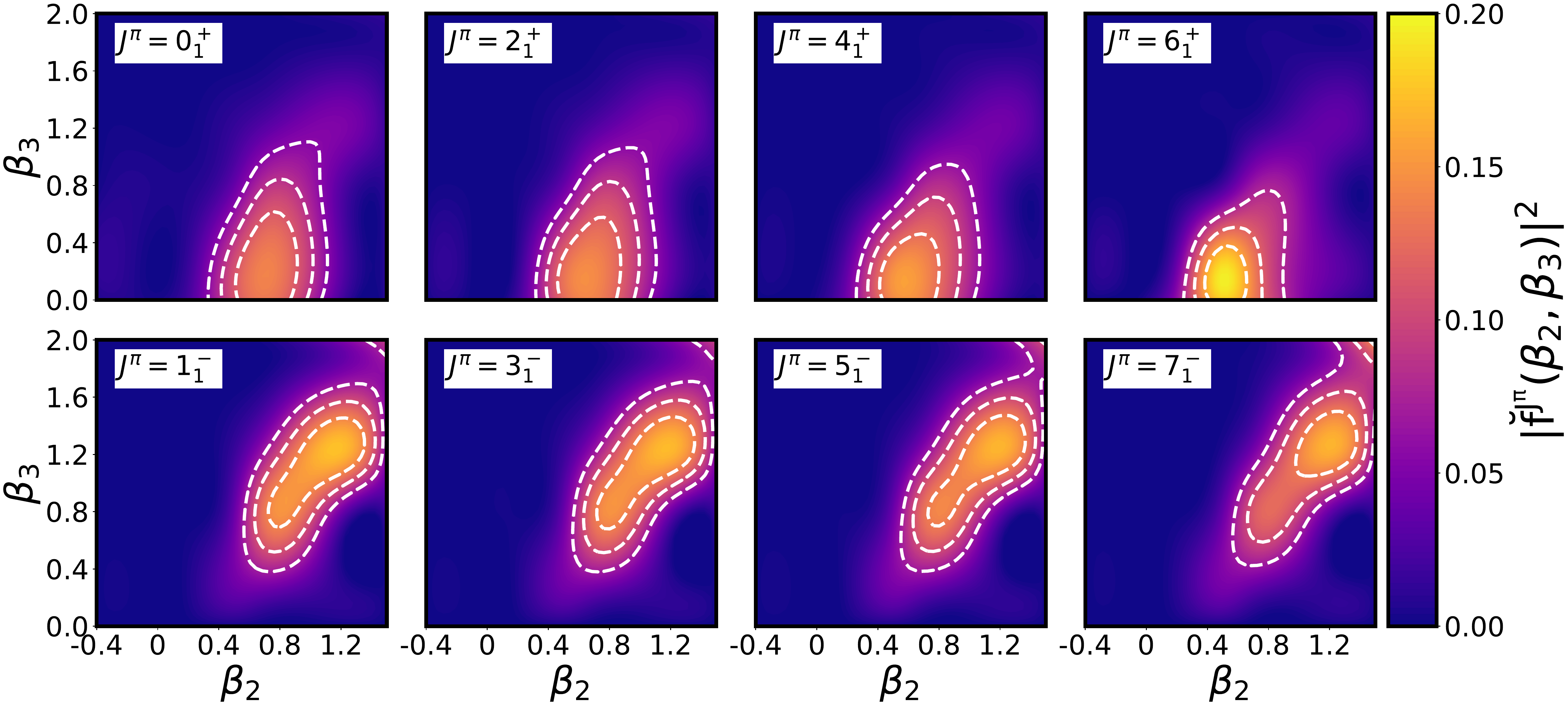}
    \caption{(Color online) Collective PGCM wave-functions in the axial $(\beta_2, \beta_3)$ plane of low-lying positive- and negative-parity states. Calculations employ the N$^3$LO $\chi$EFT Hamiltonian with $\lambda_{\text{srg}}=1.88 $\,fm$^{-1}$.}
    \label{fig:collectiveWFNe20}
\end{figure*}

The above results are qualitatively very similar to those obtained through relativistic multi-reference energy density functional (MR-EDF) calculations~\cite{zhou16a,mar18}. While being also quantitatively close, the present ab initio calculation produces more rigid TES than the EDF ones, especially on the oblate side.  

\subsubsection{Low-lying spectroscopy}

Based on the PHFB states associated with the grid displayed in Fig.~\ref{fig:HFBTES}, a PGCM calculation of $^{20}$Ne is performed. 

Figure~\ref{fig:collectiveWFNe20} displays the collective wave functions $\{\breve{f}^{\tilde{\sigma}}_{\mu}(q); q \in \text{set}\}$ (see App.~\ref{linear-redund_HWG}) of low-lying yrast states with positive and negative parities. Along the positive-parity band, collective wave-functions are peaked around a reflection-symmetric prolate configuration located at ($\beta_2 =0.55$, $\beta_3 =0$). While displaying significant shape fluctuations, in particular along the octupole degree of freedom as expected from the TES, the collective wave functions become more concentrated with increasing angular momentum, thus indicating a stabilization of the nuclear shape under rotation. The behavior is different along the negative-parity band given that the wave-functions extend over a larger range of deformations that does not decrease with $J$. Overall, one observes a significant contribution of reflection-asymmetric shapes along the positive-parity and the presence of a negative-parity band at low energy built on an octupole vibration as was anticipated from the HFB TES.

\begin{figure*}
    \centering
    \includegraphics[width=\textwidth]{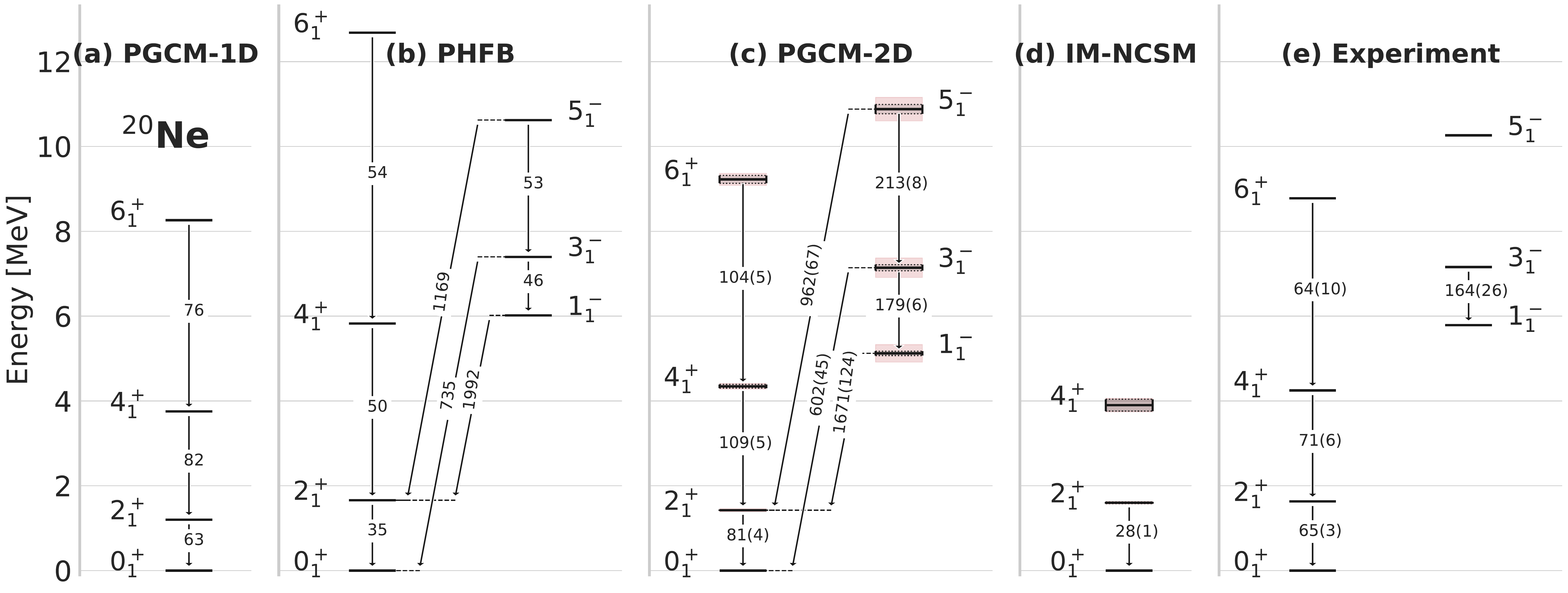}
    \caption{(Color online) Low-lying positive- and negative-parity bands in $^{20}$Ne. The intra-band $E2$ transition strengths (in e$^2$fm$^4$) are indicated along vertical arrows whereas a selection of $E3$ transition strengths (in e$^3$fm$^6$) are indicated along oblique lines. Panel (a): PGCM results obtained by restricting the mixing to the quadrupole axial degree of freedom. Panel (b): PHFB results based on the HFB configuration corresponding to the minimum of the $0^{+}$ TES located  at ($\beta_2 =0.75$, $\beta_3 =0.53$) (see Fig.~\ref{fig:PHFBTES}). Panel (c): PGCM results obtained using the set of points in the axial $(\beta_2, \beta_3)$ plane displayed in Fig.~\ref{fig:HFBTES}. Panel (d): IM-NCSM results. Panel (e): experimental data. PGCM results in panel (c) display model-space (black box) plus  $\chi$EFT (pink band) uncertainties.  IM-NCSM results in panel (d) display total many-body (black box) plus $\chi$EFT (pink band) uncertainties. The N$^3$LO $\chi$EFT Hamiltonian with $\lambda_{\text{srg}}=1.88 $\,fm$^{-1}$ is employed in PGCM and IM-NCSM calculations.}
    \label{fig:spectreNe20}
\end{figure*}

The low-lying spectrum corresponding to the collective wave-functions displayed in Fig.~\ref{fig:collectiveWFNe20}, labeled PGCM-2D, is compared in Fig.~\ref{fig:spectreNe20} (panel (c)) to experimental data (panel (e)) and to IM-NCSM results (panel (d)). Experimental excitation energies are consistently reproduced by PGCM and IM-NCSM results. IM-NCSM results, which act as quasi-exact solutions for the employed Hamiltonian, are thus reproduced by the 2D axial GCM within their respective uncertainties. As explained earlier, the rank-reduction of the three-nucleon force (see Sec.~\ref{uncertainPGCM}) and missing dynamical correlations (see Paper III) would add several percents of uncertainties to the PGCM results, making both sets of results fully compatible. One notices that the $\chi$EFT uncertainty at N$^3$LO is estimated to be sub-leading compared to many-body uncertainties in both sets of calculations. The agreement between both theoretical spectra is remarkable given that individual PGCM energies are about $60$\,MeV away from the converged values due to missing dynamical correlations (see Sec.~\ref{energiesNechain} along with Paper III for a detailed discussion). This proves that dynamical correlations contribute (essentially) identically to the energy of all low-lying states whereas static correlations are essential to describe their (mostly) collective nature. While constituting the mere first order of the PGCM-PT expansion, the PGCM is thus, at least in the present example, well suited in itself to describe the low-lying spectrum. Still, one expects dynamical correlations to provide sub-leading corrections.

\begin{figure*}
    \centering
    \includegraphics[width=1\textwidth]{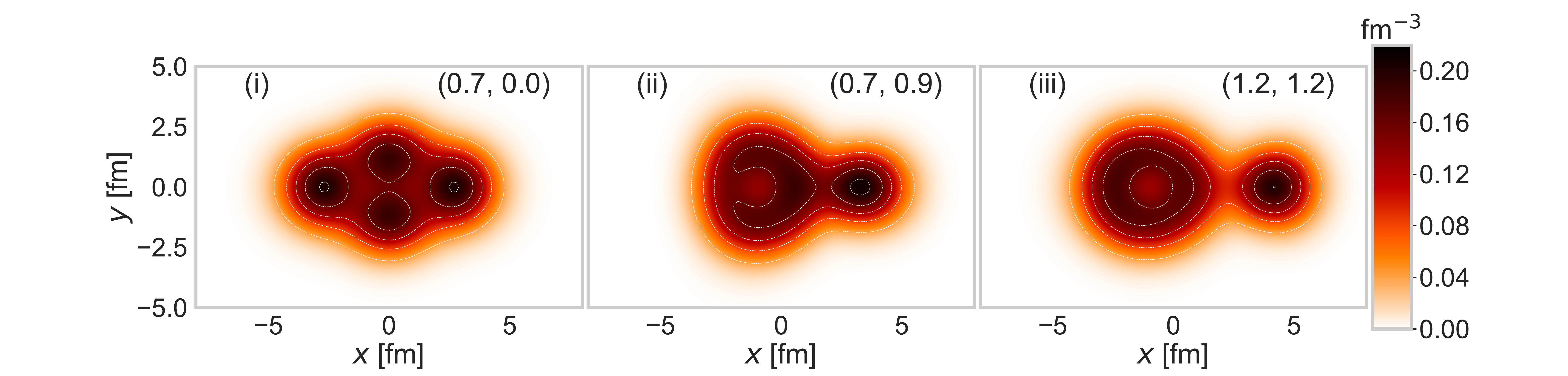}
    \caption{(Color online) Point matter distribution of $^{20}$Ne in the x-y plane corresponding to three constrained HFB configurations located at (i) ($\beta_2 =0.7$, $\beta_3 =0$), (ii) ($\beta_2 =0.7$, $\beta_3 =0.9$) and (iii) ($\beta_2 =1.2$, $\beta_3 =1.2$) in the axial ($\beta_2$, $\beta_3$) plane. Calculations employ the N$^3$LO $\chi$EFT Hamiltonian with $\lambda_{\text{srg}}=1.88$\,fm$^{-1}$.}
    \label{fig:intrinsincdensNe20}
\end{figure*}

Electric quadrupole transition strengths within each band are decently accounted for by the PGCM calculation, although being too large by a factor 1.1-1.6 compared to experimental data and by a factor 2.9 compared to the reference IM-NCSM $B(E2: 2_1^{+}\rightarrow 0_1^{+})$ value\footnote{While IM-NCSM energies and radii are very robust, it is less clear for $B(E2)$ values at this point in time such that the reference should be taken with a grain of salt.}. One expects missing dynamical correlations to reduce the collective character of the states and thus to decrease the $B(E2)$ transitions. One also notes that relativistic MR-EDF calculations~\cite{zhou16a} produced smaller $B(E2)$ transitions by spreading the collective wave-functions onto the oblate side\footnote{Excitation energies of the positive parity band were however slightly worse than in the present calculation.},  which does not happen here due to the stiffer TES. 

Interestingly, limiting the PGCM mixing to reflection-symmetric HFB states (panel (a)) compresses too much the positive-parity band in addition to forbidding the access to the negative-parity one. Contrarily, reducing the approach to a PHFB calculation based on the sole reflection-asymmetric HFB state located at the minimum of the $0^{+}$ PHFB TES ($\beta_2 =0.75$, $\beta_3 =0.53$) spreads out the positive-parity band too much and reduces too significantly the collectivity in the negative-parity band compared to experiment (panel (b)). The 2D PGCM calculation of reference is optimal and situated in between these two limiting cases, which indicates not only the need for octupole configurations but also for their fluctuations.

\subsubsection{Density distributions}

\begin{figure}
    \centering
    \includegraphics[width=.5\textwidth]{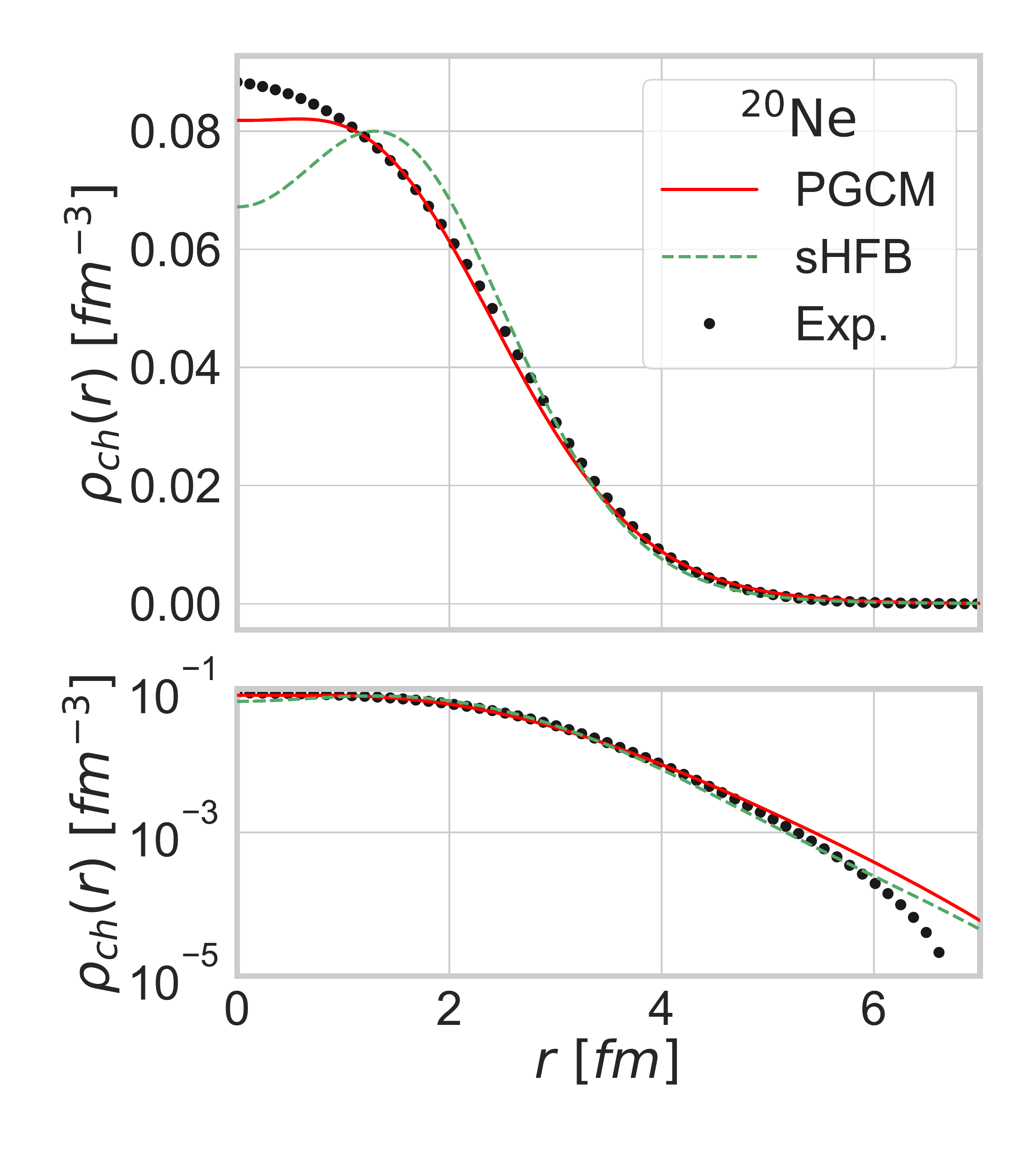}
    \caption{(Color online) Spherical HFB, PGCM and experimental $^{20}$Ne ground-state charge density distributions in linear (upper panel) and logarithmic (lower panel) scales. Calculations employ the N$^3$LO $\chi$EFT Hamiltonian with $\lambda_{\text{srg}}=1.88 $\,fm$^{-1}$.}
    \label{fig:labdensNe20}
\end{figure}

Point matter densities of $^{20}$Ne associated with three different HFB configurations are displayed in the x-y plane in Fig.~\ref{fig:intrinsincdensNe20}. The three chosen configurations correspond to (i) the maximum of the $0^+$ ground-state collective wave-function ($\beta_2 =0.7$, $\beta_3 =0$), (ii) the half-maximum of the $0^+$ ground-state collective wave-function with the largest octupole deformation  ($\beta_2 =0.7$, $\beta_3 =0.9$) and (iii) the maximum of the $1^-$ state collective wave-function ($\beta_2 =1.2$, $\beta_3 =1.2$). Panels (i) and (ii) demonstrate that the ground-state not only displays clustering but actually mixes configurations ranging from a dominant compact $\alpha+^{12}\text{C}+\alpha$ structure to a sub-leading quasi-$^{16}\text{C}+\alpha$ structure. Panel (iii) proves that the low-lying negative parity band is built out of a proper $^{16}\text{C}+\alpha$ cluster structure.

\begin{figure*}
    \centering
    \includegraphics[width=1.05\textwidth]{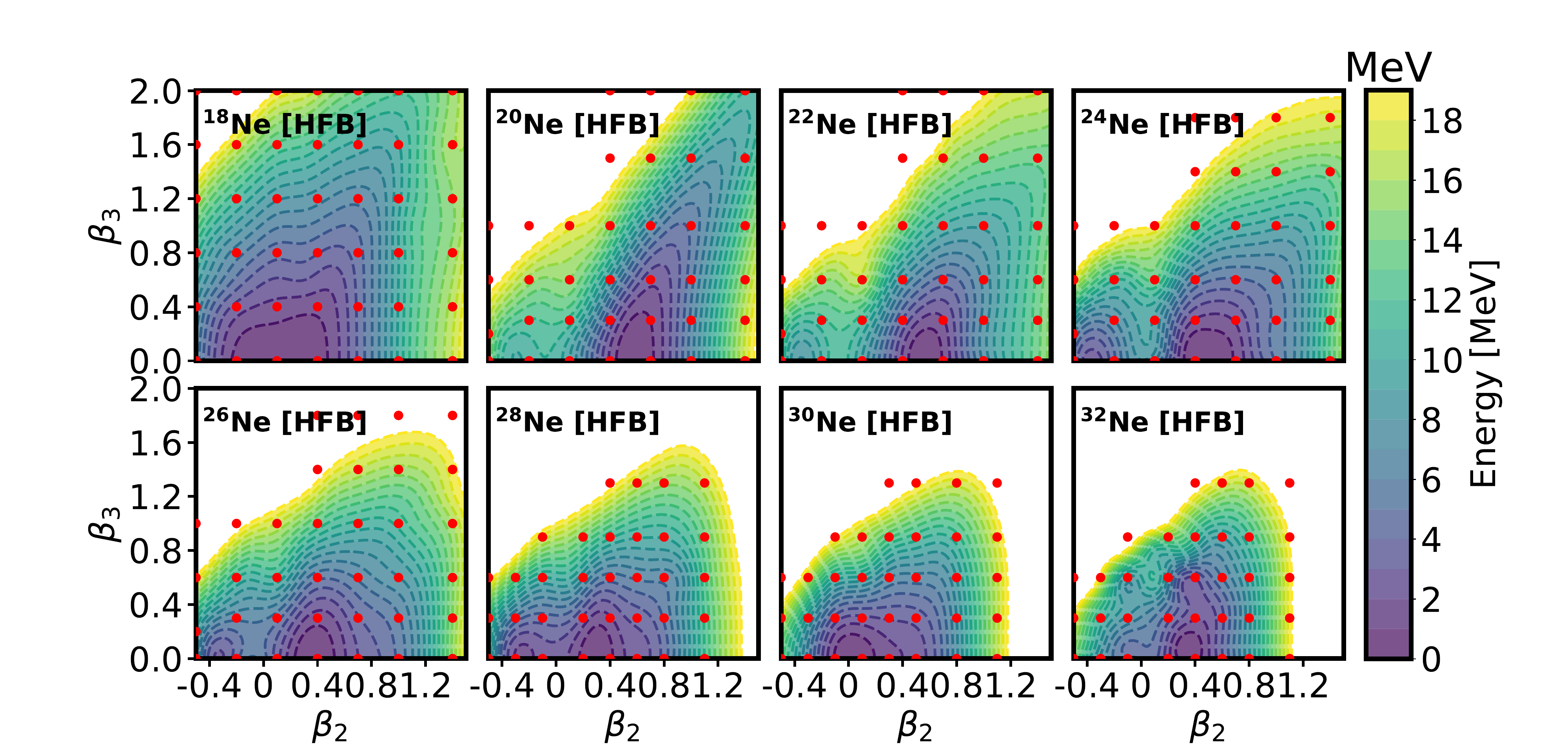}
    \caption{(Color online) HFB TES of $^{18-32}$Ne in the axial $(\beta_2, \beta_3)$ plane. The (red) dots characterize in each case the set of HFB states used in the subsequent PGCM calculations. Calculations employ the N$^3$LO $\chi$EFT Hamiltonian with $\lambda_{\text{srg}}=1.88 $\,fm$^{-1}$.}
    \label{fig:PHFBTESNechain}
\end{figure*}

\begin{figure}
    \centering
    \includegraphics[width=.5\textwidth]{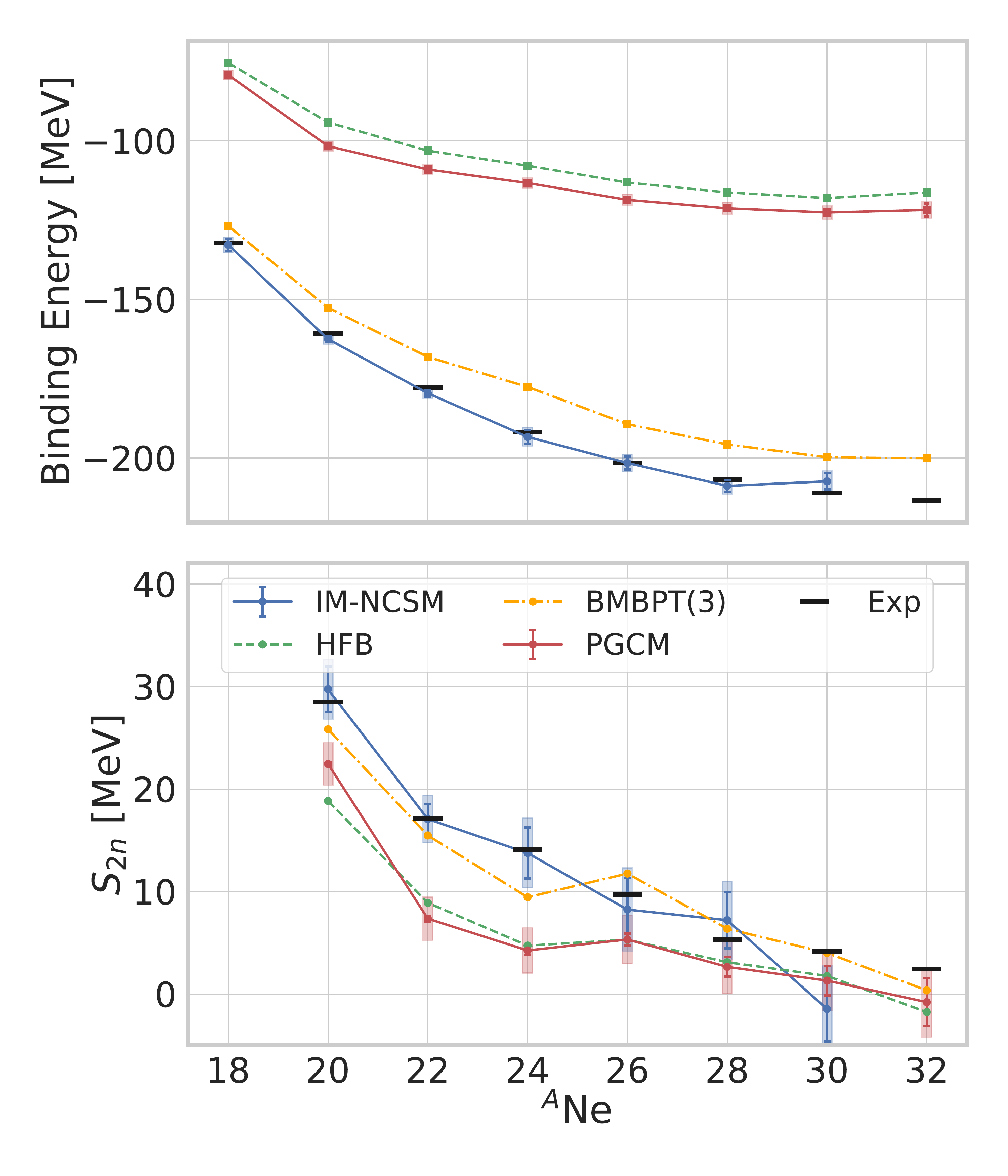}
    \caption{(Color online) Absolute ground-state energies (upper panel) and two-neutron separation energies (lower panel) along the Ne isotopic chain. Results from HFB, PGCM, BMBPT and IM-NCSM calculations are compared to experimental data. The N$^3$LO $\chi$EFT Hamiltonian with $\lambda_{\text{srg}}=1.88 $\,fm$^{-1}$  is employed in HFB/PGCM/BMBPT/IM-NCSM calculations. BMBPT calculations are performed with \(e_{\text{max}} = 10\).}
    \label{fig:GSenergiesNechain}
\end{figure}

Of course, intrinsic cluster structures are not observable per se and can only be probed indirectly. Still, the observable charge density distribution displays fingerprints of many-body correlations among which are the strong static correlations associated with intrinsic shape deformation and fluctuation. In order to illustrate this feature, the radial PGCM  charge density distribution of the $0^+$ ground-state is compared to experimental data and to the charge density computed from the spherical HFB (sHFB) configuration in Fig.~\ref{fig:labdensNe20}. Charge density distributions with respect to the center of mass are obtained from point-proton and point-neutron density distributions according to the procedure described in App.~\ref{ch_density}. As visible from the upper panel of Fig.~\ref{fig:labdensNe20}, the PGCM charge density reproduces very satisfactorily the experimental data. While it is too low in the center of the nucleus, many-body correlations partly fill up the artificial depletion displayed at the nuclear center by the sHFB density and suppress the latter accordingly in the interval $r \in [1,2]$\,fm. Furthermore, static correlations associated with shape deformation and fluctuation increase the charge density distribution in the interval $r \in [4,5]$\,fm to improve the agreement with experimental data. However, and as visible in the lower panel of Fig.~\ref{fig:labdensNe20}, the long tail part of the PGCM density overshoots the experimental density. This is consistent with both the too low two-neutron separation energy and the too high rms charge radius $r_{\text{ch}}$ discussed later on.

\subsection{Isotopic chain}

The PGCM spectroscopic results obtained in the non-trivial $^{20}$Ne isotope are very encouraging. In order to deepen the analysis, the study is now extended to other Ne isotopes and to additional observables. 

\subsubsection{Total energy surfaces}

\begin{figure*}
    \centering
    \includegraphics[width=0.9\textwidth]{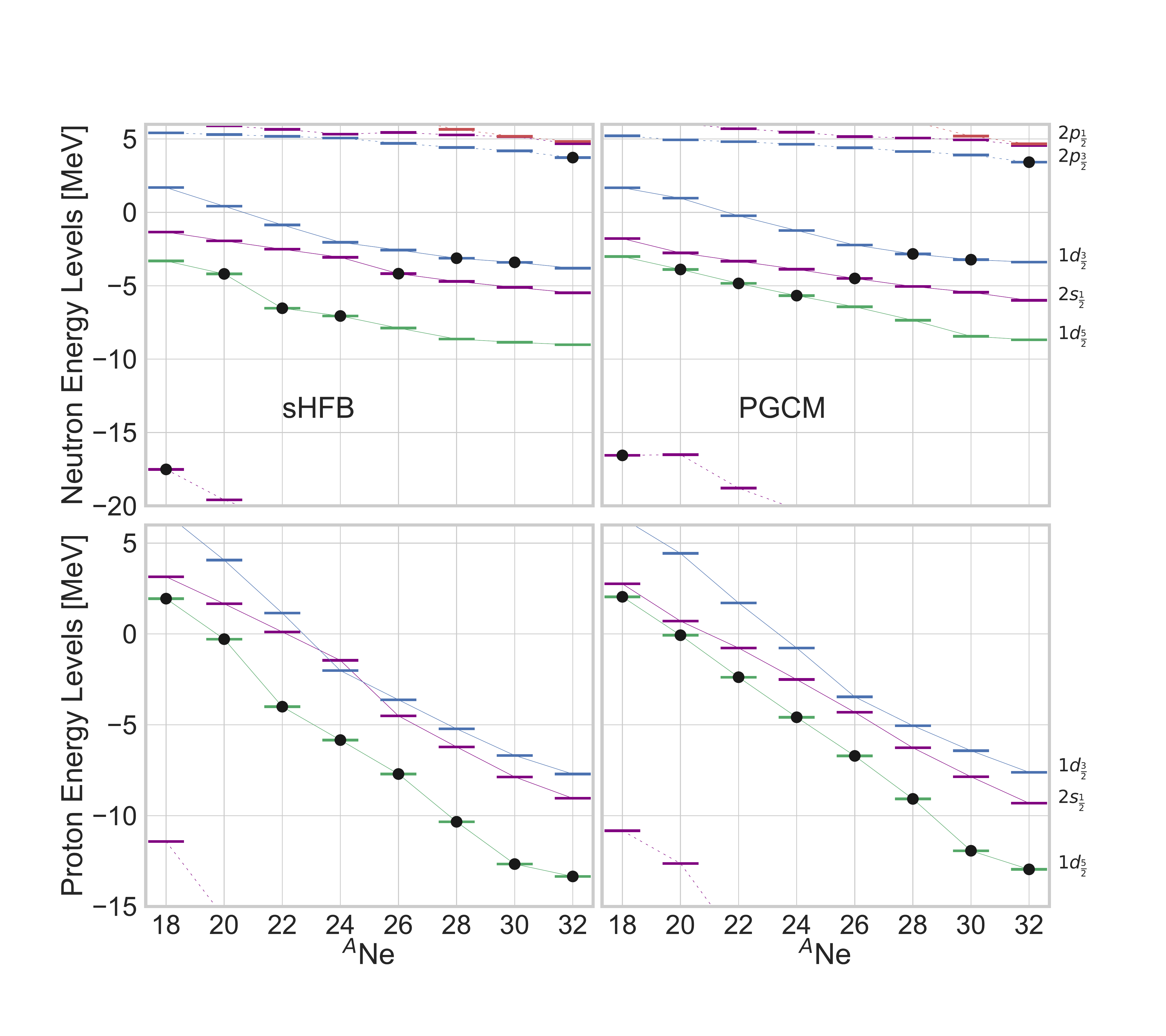}
    \caption{(Color online) Evolution of neutron (upper panels) and proton (lower panels)  Baranger's single-particle spectra of the $0^+$ ground-state in $^{18-32}$Ne. Results from spherical HFB (left column) and PGCM (right column) calculations in the axial $(\beta_2, \beta_3)$ plane are shown. Black dots denote the last occupied orbital associated with a naive filling of the shells. Calculations employ the N$^3$LO $\chi$EFT Hamiltonian with $\lambda_{\text{srg}}=1.88 $\,fm$^{-1}$.}
    \label{fig:spectreNechainBaranger}
\end{figure*}

Figure~\ref{fig:PHFBTESNechain} shows the evolution of the HFB TES in the axial $(\beta_2, \beta_3)$ plane along the Neon chain. The already studied $^{20}$Ne nucleus (Fig.~\ref{fig:HFBTES}) appears to be transitional between $^{18}$Ne, whose TES is very soft in both $\beta_2$ and $\beta_3$ directions, and heavier Ne isotopes that become increasingly rigid against octupole deformation\footnote{Octupole collectivity typically builds from strong correlations between nucleons near the Fermi surface sitting on orbitals of opposite parity and differing by three units of angular momentum $\Delta l = 3$. While such conditions are fulfilled for proton and neutron numbers $Z,N \approx 34,56,88,134$ ~\cite{but96}, one also expects a large softness against octupole deformation in light self-conjugate nuclei featuring an asymmetric di-nucleus clustering, such as the $^{16}\text{O}+\alpha$ configuration of $^{20}$Ne~\cite{hor68,mar83}. The fragmentation of the nucleus in two symmetric or asymmetric clusters can be understood from the dynamical symmetries of the anisotropic harmonic oscillator potential with frequencies in rational ratios~\cite{naz92}. Super-deformed systems are susceptible to cluster into two asymmetric (symmetric) spherical fragments for proton and neutron numbers  $Z,N \approx 2,10,28,60,110,182$ ($Z,N \approx 4,16,40,80,140$). Consequently, Ne isotopes with neutron numbers close to 8 are expected to be soft against octupole deformation, while a competition between the development of octupole collectivity (due to protons) and the restoring force towards a reflection-symmetric configuration (due to neutrons) makes neutron-rich Ne isotopes stiffer against octupole deformations. The interpretation of these features in terms of molecular covalent bonds is developed in Ref.~\cite{mar18}.}. The quadrupole deformation of the prolate minimum decreases gradually to reach the spherical $^{30}$Ne isotope before increasing again in $^{32}$Ne. At the same time, the softness against quadrupole deformation fluctuates, as the absolute prolate minimum becomes connected to a local oblate minimum in $^{24-28}$Ne before spreading again on the prolate side in $^{30}$Ne to generate a non-zero prolate deformation in $^{32}$Ne.

\subsubsection{Ground-state energies}
\label{energiesNechain}

The upper panel of Fig.~\ref{fig:GSenergiesNechain} displays the absolute ground-state energy along the Ne isotopic chain. Experimental data are well reproduced by IM-NCSM calculations within uncertainties, with a slight anomaly at $^{30}$Ne, which seems to be less bound than $^{28}$Ne based on the central value. 
Overall, the results are similar, and even better, than coupled cluster calculations in the SDT-1 approximation (CCSDT-1) performed with the $\Delta$NNLO$_{\text{GO}}$(394) Hamiltonian~\cite{Novario:2020kuf}. Strikingly, PGCM binding energies miss between $60$ and $90$\,MeVs, an underbinding that increases with neutron number. The gain compared to HFB energies is small on that scale (i.e. $\sim5-7$\,MeV) and could never compensate, even with a more elaborate PGCM ansatz, for the difference that is obviously due to missing dynamical correlations. While the goal is to bring in these correlations within a symmetry-conserving scheme, i.e. on top of the PGCM unperturbed state via PGCM-PT~\cite{frosini21b}, their effect can already be appreciated through the results of single-reference Bogoliubov many-body perturbation theory (BMBPT)~\cite{Duguet:2015yle,Tichai18BMBPT,Arthuis:2018yoo,Demol:2020mzd,Tichai2020review} calculations performed on top of a deformed HFB reference state\footnote{First results of this kind were presented in Ref.~\cite{Frosini:2021tuj}.} that are displayed in Fig.~\ref{fig:GSenergiesNechain}. The bulk of correlations is indeed recovered at the BMBPT(3) level, and we note that the gain in energy increases with neutron number and therefore corrects the overall trend at the same time. BMBPT(3) energies are still about $7-15$\,MeV away from IM-NCSM and experimental values, which is similar in magnitude to the static correlations gained via symmetry restorations and shape fluctuations within the PGCM\footnote{In $^{20}$Ne, one has $E_{\text{BMBPT}}=152.6$\,MeV, $E_{\text{IM-NCSM}}=162.6$\,MeV and $E_{\text{PGCM}}-E_{\text{HFB}}=7.4$\,MeV, knowing that $E_{\text{EXP}}=160.6$\,MeV. It must be noted that, just as BMBPT, CCSDT-1 calculations relying on a purely "vertical" expansion on top of a deformed mean-field state also provides slightly underbound Ne isotopes with the $\Delta$NNLO$_{\text{GO}}$(394) Hamiltonian~\cite{Novario:2020kuf} and thus require the addition of 3-5 MeVs of static correlations associated with symmetry restoration and shape fluctuations.}. Thus the {\it consistent} ``sum'' of static and dynamical correlations accessible via PGCM-PT can be expected to bring the absolute values very close to IM-NCSM results; see Paper III for a related discussion. 

The lower panel of Fig.~\ref{fig:GSenergiesNechain} displays two-neutron separation energies $S_{2n}$ to appreciate the stability of Ne isotopes against two-neutron emission. The IM-NCSM results reproduce experimental data well within uncertainties, with the exception of $^{30}$Ne, which shows the aforementioned anomaly leading to a slightly negative central value with a sizable error bar that almost overlaps with  experiment. Consistently with the too flat curve in the upper panel, PGCM $S_{2n}$ are too low across the chain such that the drip-line is wrongly predicted to be located at $^{30}$Ne instead of $^{34}$Ne~\cite{Ahn19}. While static collective correlations captured through PGCM have no impact on the $S_{2n}$, the comparison with IM-NCSM (or CCSDT-1) results underlines the importance of dynamical correlations to reproduce the evolution of binding energies with neutron number. As a matter of fact, dynamical correlations brought in at the BMBPT(3) level correct for the wrong trend of HFB binding energies such that the $S_{2n}$s become perfectly consistent with IM-NCSM results and experimental data. Once again, there is no obvious reason to believe that consistently correcting PGCM results for dynamical correlations will not bring the same benefit.

\begin{figure}
    \centering
    \includegraphics[width=0.5\textwidth]{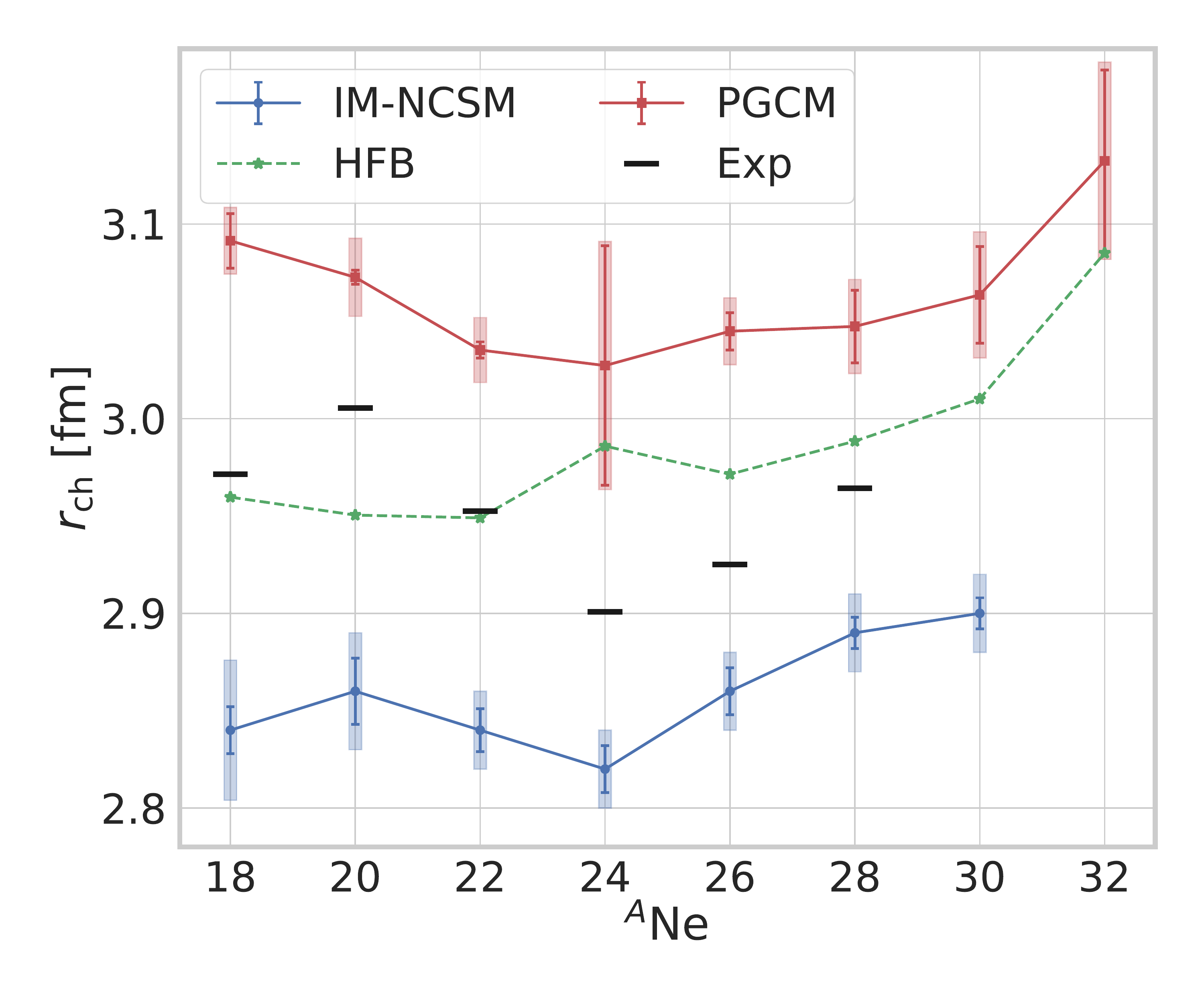}
    \caption{(Color online) Theoretical (HFB, PGCM, BMBPT, IM-NCSM) and experimental ground-state rms charge radius along the Neon isotopic chain. HFB and BMBPT results correspond to the HFB minimum in the axial $(\beta_2, \beta_3)$ plane. PGCM calculations are performed over the axial $(\beta_2, \beta_3)$ plane. The N$^3$LO $\chi$EFT Hamiltonian with $\lambda_{\text{srg}}=1.88 $\,fm$^{-1}$  is employed in PGCM and IM-NCSM calculations. BMBPT calculations are performed with \(e_{\text{max}} = 10\).}
    \label{fig:rchNechain}
\end{figure}

To further put the binding energy evolution in perspective within our theoretical scheme, Fig.~\ref{fig:spectreNechainBaranger} displays the evolution of neutron and proton (non-observable) Baranger's spherical shell structure~\cite{Duguet:2011sq,Duguet:2014tua} along the Neon chain for both spherical HFB and PGCM $0^+$ ground states\footnote{Baranger's single-particle energies embody the genuine one-body shell structure that can be extracted from any many-body calculation~\cite{Duguet:2011sq,Duguet:2014tua}, i.e. their definition is not associated with a mean-field approximation as the HF single-particle energies are for example.}. The last occupied orbit associated with a naive filling of the shells is indicated with a black dot for each isotope. One first observes that static correlations do tend to compress the spectrum around the Fermi energy but without changing it qualitatively here. The most important feature for the present discussion relates to the very large gap between neutron sd and pf shells. The fact that this gap is barely compressed going from sHFB to PGCM such that the neutron 2p$_{3/2}$ remains unbound demonstrates that cross-shell correlations in the PGCM state are insufficient to bind $^{32,34}$Ne and probably too weak already in $^{28,30}$Ne to properly describe the physics of the island of inversion.

\subsubsection{Ground-state rms charge radii}

Figure~\ref{fig:rchNechain} displays the ground-state charge rms radius of $^{18-32}$Ne. One first observes that IM-NCSM results are systematically too low compared to experimental data by about $0.08-0.13$\,fm whereas the trend with neutron number is consistent up to $^{28}$Ne, which is the most neutron-rich isotope for which  $r_{\text{ch}}$ is known experimentally. The known sub-shell closure at $N = 14$ is nicely captured, as via CCSDT-1 calculations performed with the $\Delta$NNLO$_{\text{GO}}$(394) Hamiltonian~\cite{Novario:2020kuf}. 
We note that the IM-NCSM results for the radius show a sensitivity to $\lambda_{\text{srg}}$, i.e., going from  $\lambda_{\text{srg}}=1.88$\,fm$^{-1}$ to $\lambda_{\text{srg}}=2.23$\,fm$^{-1}$ (not shown) the charge radius increases by about $0.05$\,fm. The radius operator for these calculation has been SRG transformed consistently at the two-body level, and we do not expect induced multi-particle contributions to the radius operator to cause this difference. We will further explore this behavior, which is also evident in other nuclei \cite{Huther_2020}, in a forthcoming publication.  

Except in $^{18-22}$Ne where octupole fluctuations are important and in $^{24}$Ne, radii associated with the HFB minimum in the axial $(\beta_2, \beta_3)$ plane roughly follow the trend of IM-NCSM predictions but are about $0.1$\,fm larger. From a phenomenological standpoint, it seems consistent that the deficit of binding at the HFB level correlates with too large radii. However, this correlation is not effective when adding static correlations via PGCM. Indeed, while increasing the binding energy by few MeVs and leaving $S_{2n}$ essentially untouched, PGCM systematically increases rms charge radii compared to HFB by mixing in more deformed configurations than the HFB minimum (see Figs.~\ref{fig:HFBTES} and~\ref{fig:collectiveWFNe20}). Eventually, PGCM results overestimate experiment (IM-NCSM) by about $0.1$\,fm ($0.3$\,fm) all throughout the isotopic chain even though the isotopic dependence is closer to IM-NCSM than HFB. Thus, static collective correlations make PGCM largely exaggerate rms charge radii and must be compensated for by missing dynamical correlations. Given that dynamical correlations directly brought on top of the deformed mean-field increase charge radii~\cite{gauteperso}, it will be of interest to see how and why they decrease charge radii when brought on top of the PGCM state via PGCM-PT.

\subsubsection{Low-lying spectroscopy}
\label{spectroNechain}

\begin{figure}
    \centering
    \includegraphics[width=0.5\textwidth]{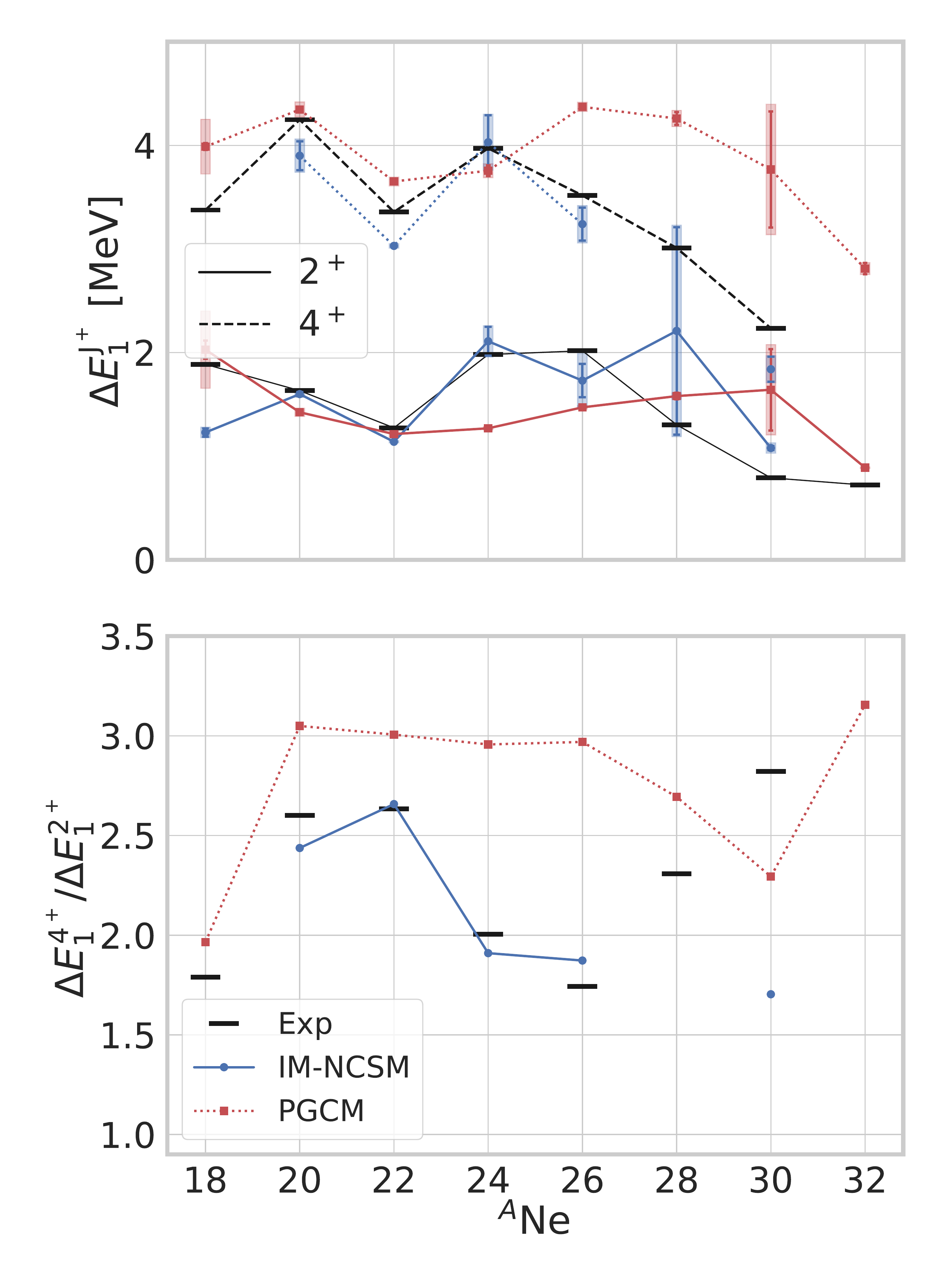}
    \caption{(Color online) Low-lying spectroscopy in $^{18-32}$Ne. First $\Delta E^{2^+}_1$ and $\Delta E^{4^+}_1$ excitation energies (upper panel) and their ratio $\Delta E^{4^+}_1/\Delta E^{2^+}_1$ (lower panel). PGCM results with model-space (black box) plus  $\chi$EFT (pink band) uncertainties and IM-NCSM results with total many-body (black box) plus $\chi$EFT (pink band) uncertainties are compared to experimental data. The N$^3$LO $\chi$EFT Hamiltonian with $\lambda_{\text{srg}}=1.88 $\,fm$^{-1}$  is employed in PGCM and IM-NCSM calculations.}
    \label{fig:spectreNechain}
\end{figure}

Figure~\ref{fig:spectreNechain} displays the systematic of the first $2^+$ and $4^+$ excitation energies in $^{18-32}$Ne. Except for the rotational character of the ground-state band in $^{30}$Ne, experimental data are well reproduced by IM-NSCM calculations all along the isotopic chain. As for PGCM calculations, the excellent results obtained in $^{20}$Ne do extend to $^{22}$Ne. Starting with $^{24}$Ne, the trend of PGCM results is however at odds with IM-NCSM and experimental values. In particular, the steep decrease of the first $2^+$ ($4^+$) energy beyond $^{26}$Ne  ($^{24}$Ne), well captured by IM-NCSM calculations, is absent from the PGCM results. Furthermore, the experimental $\Delta E^{4^+}_1/\Delta E^{2^+}_1$ ratio displayed in the lower panel of Fig.~\ref{fig:spectreNechain} demonstrates that the nature of the ground-state band changes brutally beyond $^{26}$Ne as one enters the island of inversion to approach the rotational value in $^{30}$Ne. As could have been anticipated from the evolution of the HFB TES in Fig.~\ref{fig:PHFBTESNechain}, this qualitative change is not captured by PGCM calculations that predict $^{30}$Ne ground-state to be spherical. 

\begin{figure}
    \centering
    \includegraphics[width=0.5\textwidth]{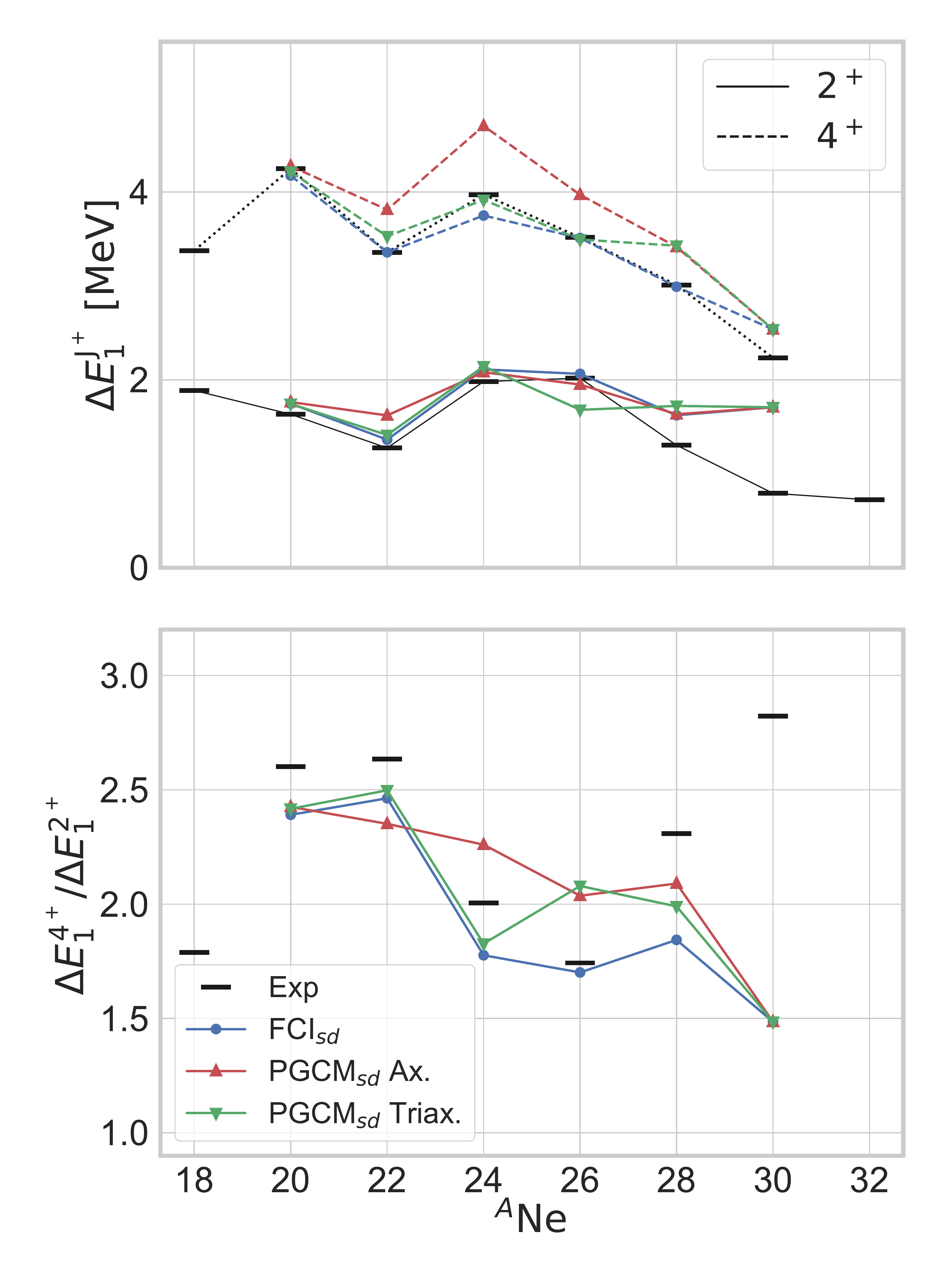}
    \caption{(Color online) Low-lying spectroscopy in $^{18-32}$Ne from sd valence-space calculations. First $\Delta E^{2^+}_1$ and $\Delta E^{4^+}_1$ excitation energies (upper panel) and their ratio $\Delta E^{4^+}_1/\Delta E^{2^+}_1$ (lower panel). PGCM results obtained by mixing states along the axial quadrupole coordinate are compared to experimental data and to full configuration interaction (FCI) results as well as to PGCM results obtained by adding the triaxial quadrupole coordinate.}
    \label{fig:spectreNechainVS}
\end{figure}

The above discussion underlines that the ability of the PGCM to nicely reproduce the low-lying spectroscopy of $^{20-22}$Ne cannot be naively and automatically generalized to all nuclei. It is however unclear whether what is observed along the Ne isotopic chain constitutes an intrinsic limitation of the PGCM ansatz, in which case all defects must be corrected by the addition of dynamical correlations, or whether a richer PGCM ansatz could already change the situation at play.

To investigate this question, the results of sd valence-space calculations  performed with the USDB interaction~\cite{brown06a} are reported on in Fig.~\ref{fig:spectreNechainVS}. The restriction to a valence space is meant to effectively remove, or largely suppress, the explicit role of dynamical correlations and see if enriching the PGCM ansatz is sufficient to reach full configuration interaction (FCI) results~\cite{Sanchez-Fernandez:2021nfg}. To do so, the PGCM on purely axial states discussed above is enriched via the explicit addition of triaxially-deformed HFB states. When restricted to a small valence space around the Fermi level, the PGCM based on axial states is able to track the exact FCI $2^+_1$ excitation energy very closely all the way to the border of the sd shell, i.e. up to $^{30}$Ne. However, the reproduction of the $4^+_1$ excitation energy is already quite off in the middle of the shell such that the $\Delta E^{4^+}_1/\Delta E^{2^+}_1$ ratio does not track the steep decrease visible in both the data and the FCI results. Enlarging the PGCM ansatz to include triaxial states significantly improves the situation up to $^{24}$Ne, in particular by capturing the drop of the ratio that is not described correctly in the PGCM ab initio calculations (see Fig.~\ref{fig:spectreNechain}). Therefore, enriching the PGCM ansatz itself helps significantly when implicitly accounting for dynamical correlations via an effective valence-space Hamiltonian. However, one has to see if this remains true in ab initio calculations where dynamical correlations are only added a posteriori, e.g., in perturbation. Beyond $^{26}$Ne both sets of PGCM calculations reproduce FCI results very well, which is somewhat anecdotal given the smallness of the configuration space as one reaches the end the sd shell and given that sd-shell valence space calculations based on traditional empirical interactions do not reproduce the physics of the island of inversion anyway. 

\begin{figure*}
    \centering
    \includegraphics[width=0.9\textwidth]{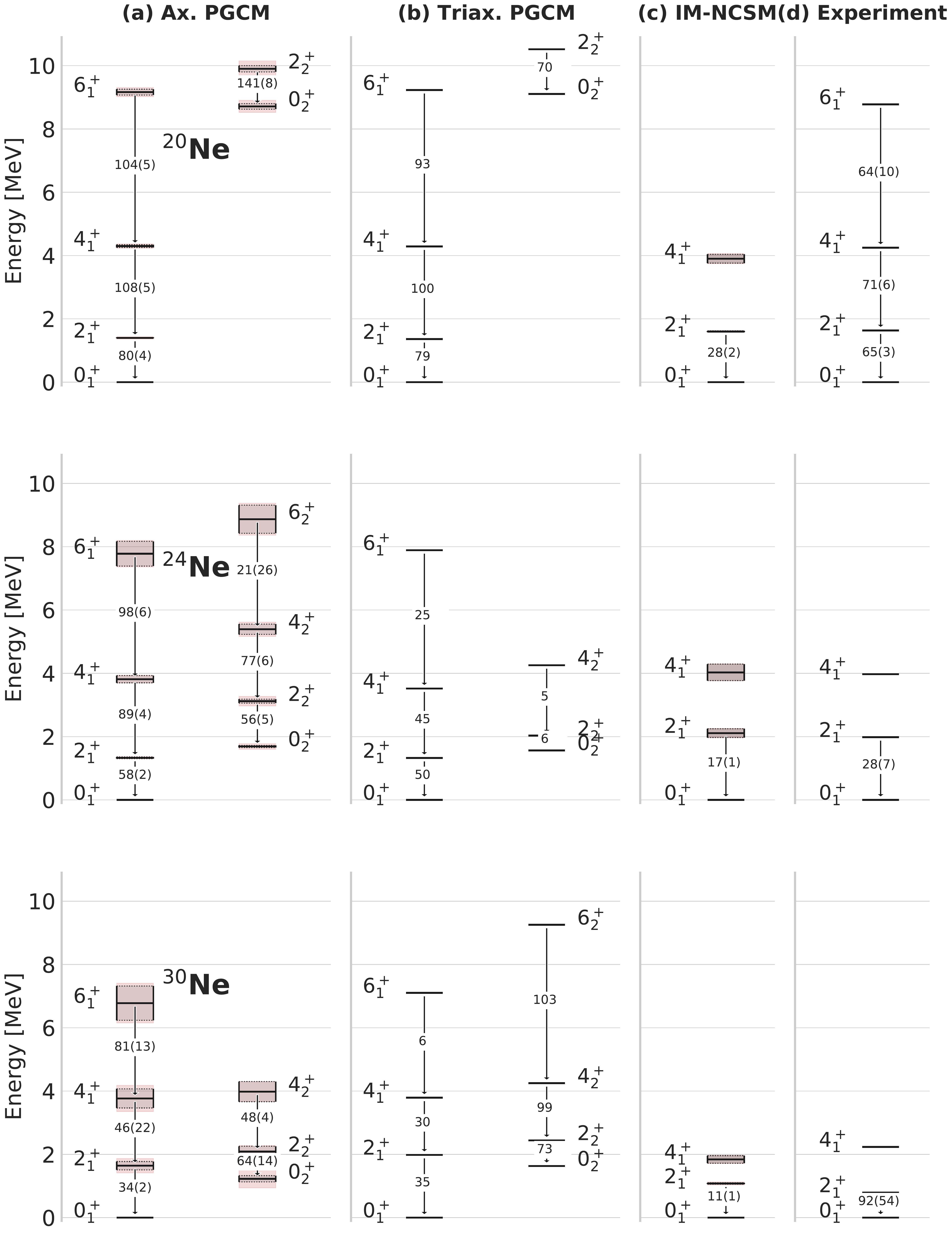}
    \caption{(Color online) First two positive-parity bands in $^{20,24,30}$Ne. The $E2$ transition strengths (in e$^2$fm$^4$) are indicated along vertical arrows. Panel (a): PGCM results obtained using HFB configurations in the axial $(\beta_2, \beta_3)$ plane. Panel (b): PGCM results obtained adding triaxially deformed HFB configurations. Panel (c): IM-NCSM results. Panel (d): experimental data.  PGCM results in panel (a) display model-space (black box) plus  $\chi$EFT (pink band) uncertainties.  IM-NCSM results in panel (c) display total many-body (black box) plus $\chi$EFT (pink band) uncertainties. The N$^3$LO $\chi$EFT Hamiltonian with $\lambda_{\text{srg}}=1.88 $\,fm$^{-1}$  is employed in PGCM and IM-NCSM calculations.}
    \label{fig:spectreNe202430triax}
\end{figure*}

\begin{figure}
    \centering
    \includegraphics[width=0.48\textwidth]{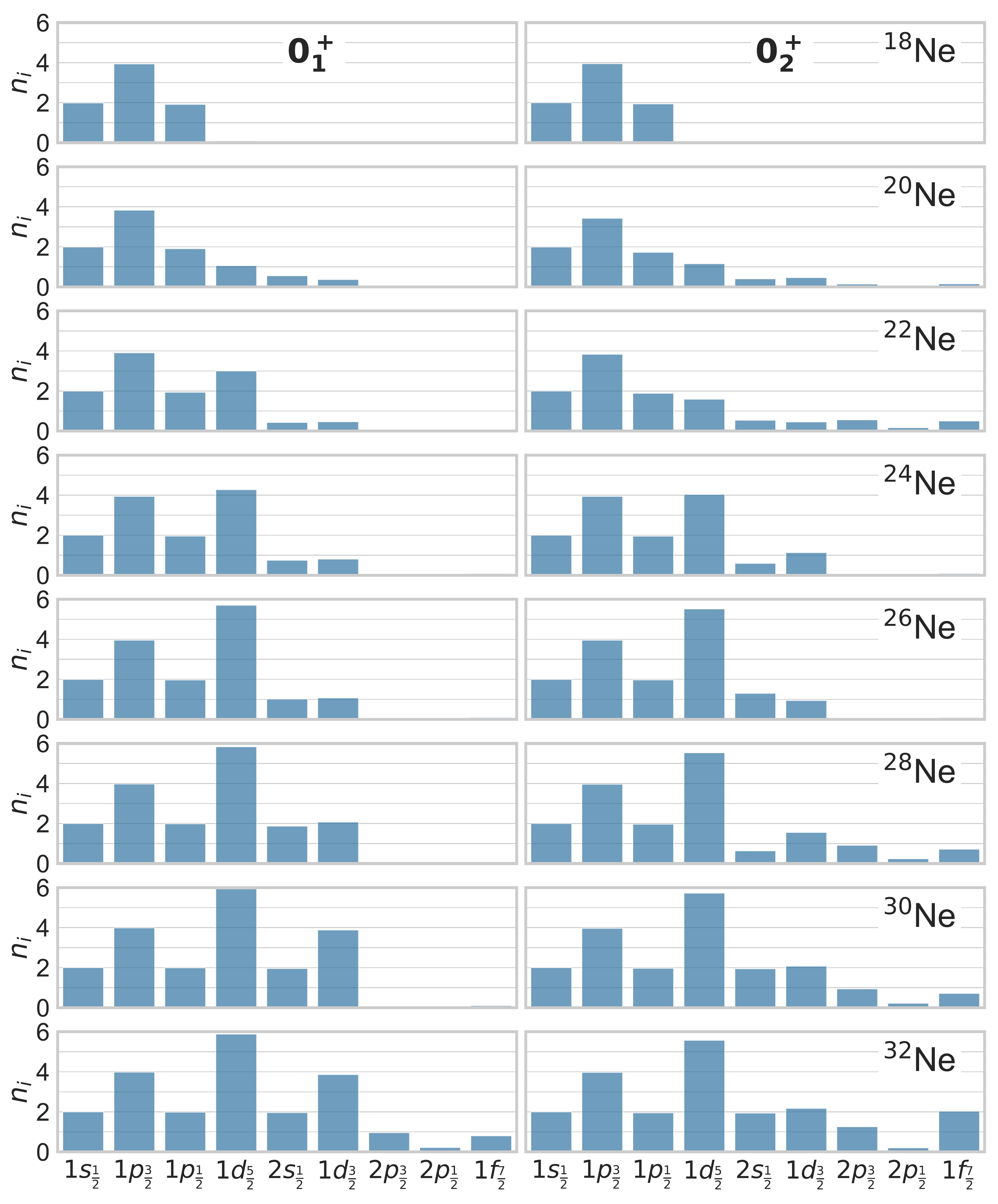}
    \caption{(Color online) Average neutron natural orbital occupations (including the $2j+1$ degeneracy factor) for the first two PGCM $0^+$ states of $^{18-32}$Ne. Calculations employ the N$^3$LO $\chi$EFT Hamiltonian with $\lambda_{\text{srg}}=1.88 $\,fm$^{-1}$.}
    \label{fig:neutronnatorbocc}
\end{figure}

Having learnt that adding triaxiality can improve the low-lying spectroscopy of certain isotopes, ab initio PGCM calculations of $^{20,24,30}$Ne have been extended to include this degree of freedom into the PGCM ansatz. Corresponding results are shown in Fig.~\ref{fig:spectreNe202430triax}\footnote{Because triaxial PGCM calculations are much more computationally intensive than axial ones, due to the two additional integrals appearing in the angular-momentum projection, associated error bars have not been computed in the present work.}. 

Starting with $^{20}$Ne, one observes that the already well reproduced spectroscopy is not spoiled by the addition of triaxial configurations. More specifically, excitation energies are barely modified whereas $B(E2)$ transition strengths within both bands are decreased, in a way that is more consistent with experimental data for the ground-state band.

In the valence-space results presented in Fig.~\ref{fig:spectreNechainVS}, the addition of triaxial configurations were mostly useful to improve the behavior of the  $\Delta E^{4^+}_1/\Delta E^{2^+}_1$ ratio around mid shell. In the ab initio calculation, the ground-state band is marginally compressed in $^{24}$Ne, moving it slightly away from IM-NCSM results and experimental data. Correspondingly, the $\Delta E^{4^+}_1/\Delta E^{2^+}_1$ ratio is only marginally lowered from $2.78$ in the axial calculation to $2.67$ in the trixial one, i.e the inclusion of the trixial degree of freedom does not bring the improvement that could be expected from valence-space calculations.

Let us now come to the most challenging $^{30}$Ne isotope located in the island of inversion. As for the axial PGCM calculation, one first notices that the rotational behavior of the experimental ground-state band is not reproduced by the IM-NCSM calculation. Thus, one cannot exclude that the computed band is not the correct one, especially given that the trend of IM-NCSM ground-state binding energies precisely presents a glitch in $^{30}$Ne as observed earlier in Fig.~\ref{fig:GSenergiesNechain}. As a matter of fact, the problem of the PGCM calculation relates to the fact that the excited intruder positive-parity band is probably the right candidate for the ground-state band, only that it is wrongly positioned above the spherical one. Indeed, both excitation energies with respect to the band-head and intra-band $B(E2)$ transitions are consistent with experimental data. This failure is consistent with the large gap between  sd and pf shells observed in the PGCM Baranger neutron spectrum  (Fig.~\ref{fig:spectreNechainBaranger}) that is a fingerprint of the lack of cross-shell correlations in the computed ground-state. While incorporating full dynamical correlations must correct for this defect, a more efficient strategy could consist in enriching the PGCM ansatz. As seen from the lower panel of Fig.~\ref{fig:spectreNe202430triax}, the triaxial PGCM does compress the intruder band and lower it slightly, but not nearly enough. At this point in time, one is thus left with two scenarios (i) the account of missing dynamical correlations inverses the order of the two bands or (ii) further enriching the PGCM ansatz brings it down\footnote{A preliminary study indicates that generating the Bogoliubov states via a variation after projection on particle number (VAPPN) calculation~\cite{Bally:2021kfw} does go in the right direction but it is not sufficient to invert the two bands per se.}, decreasing or even cancelling the need for dynamical correlations to operate the inversion. Still, given that the IM-NCSM ground-state band is not rotational (even within estimated uncertainties) and that the $\Delta E^{4^+}_1/\Delta E^{2^+}_1$ ratio is in fact close to the PGCM value, one must contemplate the fact that the Hamiltonian is to be blamed, i.e. that the associated uncertainties are underestimated.

Overall, the conclusion is that the further inclusion of the trixial degree of freedom does not change the situation in any decisive way in the present examples. Correspondingly, the island of inversion presents a challenging test case for ab initio calculations. 

\subsubsection{Natural orbitals average occupation}

To further analyze the results displayed above, the (non-observable) average occupation of neutron natural orbitals, i.e. the eigenvalues of the PGCM one-body density matrix, are displayed in Fig.~\ref{fig:neutronnatorbocc} for the first two $0^+$ states in $^{18-32}$Ne. The  $^{28-30}$Ne data confirm that, within the present theoretical calculation, the band built on the excited $0^+_2$ state is the intruder band benefiting, although not enough at the strict PGCM level, from correlations built out of particle-hole excitations into the pf shell. Contrarily, the pf natural orbitals display zero occupation for the $0^+_1$ ground-state in these two isotopes belonging to the island of inversion. Again, while enriching the PGCM ansatz can improve the situation, adding dynamical correlations associated with explicit particle-hole excitations into the pf shell is likely  to be necessary to make the intruder band become the ground-state one.

\begin{figure*}
    \centering
    \includegraphics[width=0.85\textwidth]{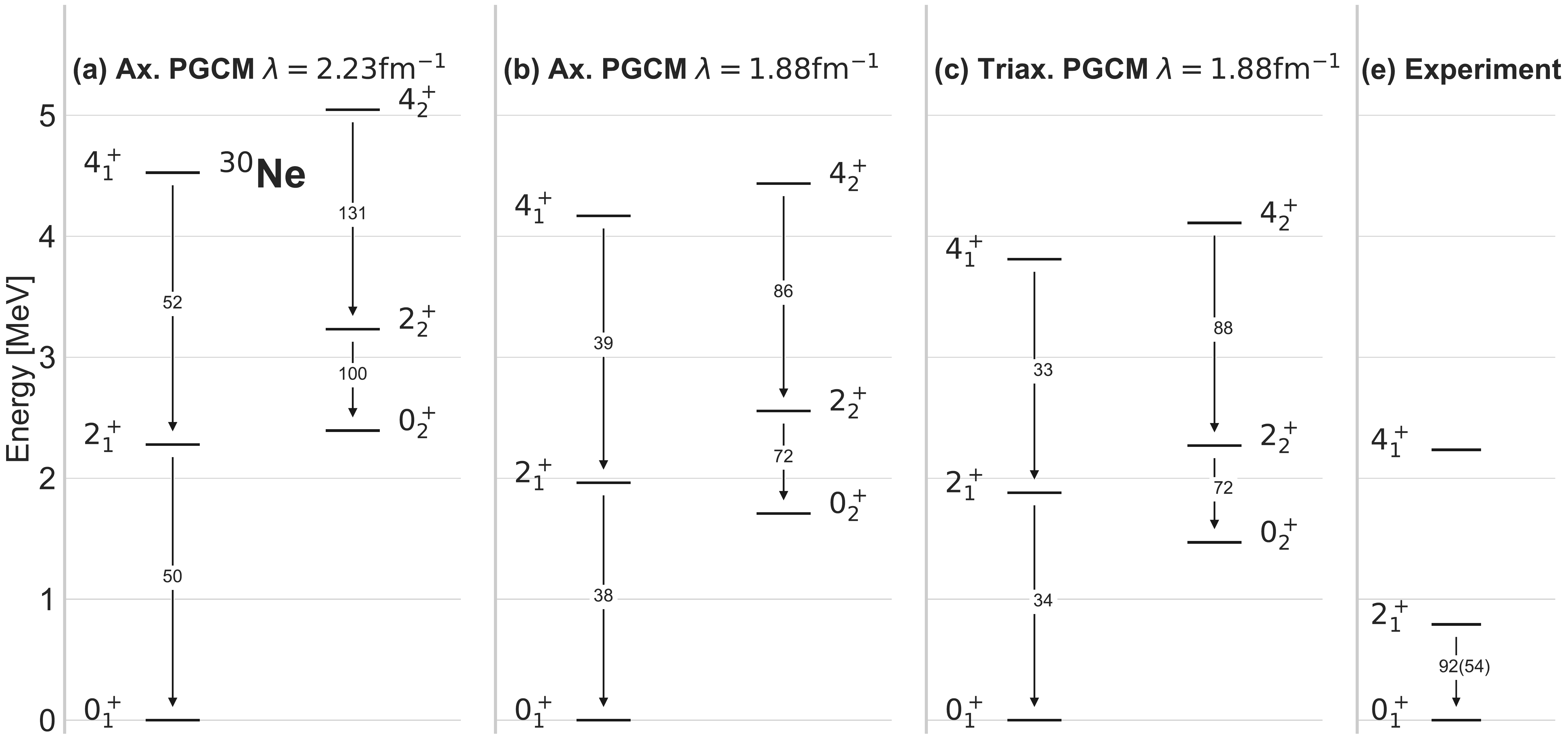}
    \caption{(Color online) Adaptation of the lower panel of Fig.~\ref{fig:spectreNe202430triax} via the addition of the axial PGCM results obtained from the N$^3$LO $\chi$EFT Hamiltonian  $\lambda_{\text{srg}}=2.23$\,fm$^{-1}$.}
    \label{fig:spectreNe30triaxsrg}
\end{figure*}

\subsubsection{SRG dependence}

The above PGCM results have been obtained from the Hamiltonian characterized by $\lambda_{\text{srg}}=1.88 $\,fm$^{-1}$. Figure~\ref{fig:spectreNe30triaxsrg} illustrates the variation of the PGCM low-lying spectroscopy in $^{30}$Ne when using  $\lambda_{\text{srg}}=2.23$\,fm$^{-1}$. Softening the Hamiltonian from $\lambda_{\text{srg}}=2.23$\,fm$^{-1}$ to $\lambda_{\text{srg}}=1.88$\,fm$^{-1}$, the PGCM spectrum is compressed, and the intruder band lowered, by as much as what was produced by the addition of  triaxial configurations.  

As visible from Tab.~\ref{tab:srddepPGCM}, $^{30}$Ne is one of the two isotopes in which the sensitivity of the spectrum is the largest, i.e. of the order of $10-20\%$. This significant dependence on  $\lambda_{\text{srg}}$ is consistent with the fact that decisive correlations associated with the island of inversion are missing in the PGCM (and IM-NCSM) calculation of this nucleus. Contrarily, the first $2^+$ and $4^+$ states in  $^{20}$Ne only vary by about $1\%$ under the same modification of $\lambda_{\text{srg}}$. Among the selected quantities, the spectroscopic quadrupole moment appears to be the most sensitive one, with variations of the order of $20\%$ throughout the isotopic chain. It will be of interest to investigate how much these variations are tamed down by the inclusion of dynamical correlations on top of PGCM.

\begin{table*}
    \centering
    \renewcommand{\arraystretch}{1.15}
    \begin{tabular}{|l|l|c|c|c|c|c|c|c|c|}
\cline{1-10}
       & &  $^{18}$Ne &  $^{20}$Ne &  $^{22}$Ne &  $^{24}$Ne &  $^{26}$Ne &  $^{28}$Ne &  $^{30}$Ne &  $^{32}$Ne \\
\cline{1-10}
$\Delta E^{J^+}_1$    & 2+ &  21.6 &   0.5 &   1.4 &   6.1 &   5.6 &  12.0 &  16.1 &   5.6 \\
       & 4+ &  23.5 &   1.5 &   0.6 &   5.3 &   1.6 &   3.2 &   8.6 &   0.4 \\
\cline{1-10}
$M1$ & 2+ &  13.6 &   0.1 &   5.9 &   1.2 &   4.2 &   0.5 &   7.2 &   1.6 \\
       & 4+ &  11.3 &   0.0 &   4.5 &   4.5 &   3.2 &   5.0 &   3.7 &   0.3 \\
\cline{1-10}
$Q2$ & 2+ &  32.6 &  20.2 &  16.6 &  21.5 &  17.8 &  19.0 &  26.8 &  16.4 \\
       & 4+ &  32.9 &  22.9 &  17.5 &  21.2 &  18.3 &  21.1 &  10.0 &  16.6 \\
\cline{1-10}
$r_{\text{ch}}$ & 0+ &   7.1 &   6.1 &   5.6 &   5.9 &   5.7 &   5.6 &   5.4 &   5.3 \\
\cline{1-10}
\end{tabular}
\caption{Percentage of variation of selected PGCM results along the Neon isotopic chain for $\lambda_{\text{srg}} \in [1.88,2.23] $\,fm$^{-1}$. Evolving the interaction systematically reduces values for radii and EM transitions.}
\label{tab:srddepPGCM}
\end{table*}

\subsubsection{Spectroscopic observables}

\begin{figure}
    \centering
    \includegraphics[width=0.5\textwidth]{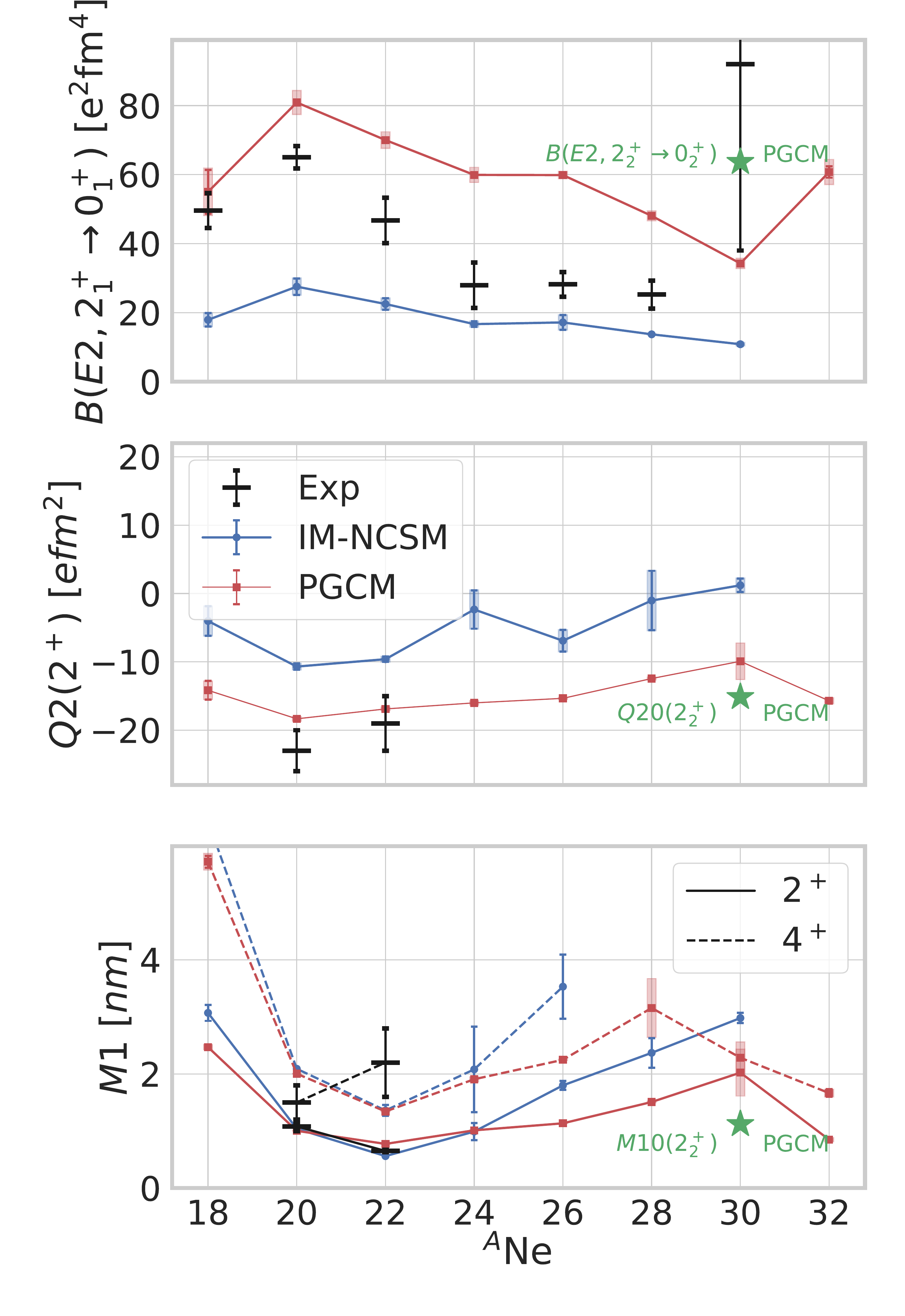}
    \caption{(Color online) Theoretical (PGCM, IM-NCSM) and experimental electromagnetic moments along the Neon isotopic chain. Upper panel: reduced electric quadrupole transition $B(E2: 2^+_1 \rightarrow 0^+_1)$ to which is added the PGCM $B(E2: 2^+_2 \rightarrow 0^+_2)$ value in $^{30}$Ne. Middle panel: spectroscopic electric quadrupole moment of the  first $2^+$ state. Lower panel: spectroscopic magnetic dipole moment of the first $2^+$ and $4^+$ states. PGCM calculations are performed in the axial $(\beta_2, \beta_3)$ plane. The N$^3$LO $\chi$EFT Hamiltonian with $\lambda_{\text{srg}}=1.88 $\,fm$^{-1}$ is employed in PGCM and IM-NCSM calculations.}
    \label{fig:Q2M1Nechain}
\end{figure}

A last set of spectroscopic quantities are displayed in Fig.~\ref{fig:Q2M1Nechain}. As visible from the upper panel, IM-NCSM $B(E2: 2^+_1 \rightarrow 0^+_1)$ transition probabilities are generally too small, particularly below $^{24}$Ne and at $^{30}$Ne even though the experimental error bars are large for this case. Contrarily, the collective character of the PGCM makes the $B(E2: 2^+_1 \rightarrow 0^+_1)$ to be well reproduced below $^{24}$Ne and overestimated between $^{24}$Ne and $^{28}$Ne. In $^{30}$Ne, the E2 transition is again too small because the first band does not correspond to the experimental one. As visible from the added point on the figure, the $B(E2: 2^+_2 \rightarrow 0^+_2)$ of the intruder band is in better agreement with experiment.

In the middle panel, the PGCM spectroscopic quadrupole moment of the first $2^+$ state reproduces well experimental data in $^{20,22}$Ne. The IM-NCSM results, however, are systematically too small in magnitude.
We have observed this type of deviation also in other cases and identified a possible explanation for this deficiency \cite{Mongelli:2021}, which is related to an enhancement of multi-particle contributions to the IMSRG-transformed quadrupole operator, which have not been included here. In other words, capturing the full collectivity in electric quadrupole observables might in some cases require multi-particle contributions to the transformed operator.  

In the lower panel, PGCM and IM-NCSM spectroscopic dipole moments of the first $2^+$ and $4^+$ states are consistent throughout the isotopic chain and nicely account for the available experimental data in $^{20,22}$Ne.

Accessing experimental electromagnetic moments in more neutron-rich Ne isotopes would allow one to better investigate the consistency of the picture that emerges from our theoretical study and would thus be welcome in the future. 

\section{Conclusions}
\label{sec_conclusions}

The second paper of the present series proposed an extensive ab initio study of neon isotopes based on in-medium no-core shell model and projected generator coordinate method calculations. The main conclusion of the present work is that, in spite of missing so-called dynamical correlations, the PGCM is shown to be a suitable ab initio method to address the low-lying spectroscopy of complex nuclei within theoretical uncertainties. For instance, the energy spectrum and electric multipole transition strengths of the low-lying parity-doublet bands in $^{20}$Ne are reproduced by taking into account the effect of octupole collective fluctuations.

Still, describing absolute binding energies, accounting consistently for a wide range of spectroscopic observables, tackling many nuclei displaying different characteristics and achieving high accuracy, eventually requires the inclusion of dynamical correlations on top of the PGCM. In fact, certain salient features, such as the physics of the island of inversion around $^{30}$Ne, require this inclusion from the outset to achieve a qualitatively correct description. This incorporation is now possible thanks to the novel multi-reference perturbation theory (PGCM-PT) formulated in the first paper of the present series~\cite{frosini21b} and that embeds the PGCM within a systematic expansion. 

The first PGCM-PT results are presented in the third paper of the present series~\cite{frosini21d}. The key question behind the present work and the associated many-body developments regards the optimal way to consistently incorporate static and dynamical correlations in view of describing complex nuclei. This is only the beginning of the journey, hence finding this optimal strategy will require time and a significant amount of trial-and-error. The third paper of the series represents a first step in this direction.

%
\begin{acknowledgements}
This project has received funding from the European Union’s Horizon 2020 research and innovation programme under the Marie Skłodowska-Curie grant agreement No. 839847. The work of TRR was supported by the Spanish MICINN under PGC2018-094583-B-I00. TM and RR were supported by the DFG Sonderforschungsbereich SFB 1245 (Project ID 279384907) and the BMBF Verbundprojekt 05P2021 (ErUM-FSP T07, Contract No. 05P21RDFNB). 

Calculations were performed in part using HPC resources from GENCI-TGCC (Contract No. A009057392) and the Lichtenberg high performance computer of the Technische Universit\"at Darmstadt. The authors also thankfully acknowledge the computer resources at Turgalium and the technical support provided by CETA-Ciemat (FI-2021-2-0013). Finally, we acknowledge the computer resources and assistance provided by GSI-Darmstadt.

\end{acknowledgements}


\begin{appendices}

\section{Linear redundancies in HWG}
\label{linear-redund_HWG}

The linear redundancies due to the  non-orthogonality of the HFB states mixed into the PGCM state must be dealt with when solving HWG's equation. Because of the manageable number of such HFB states, it can be done by diagonalizing the norm matrix ${\bold N}^{\tilde{\sigma}}$ and by removing the eigenvectors associated with eigenvalues smaller than a given threshold $\epsilon_{\text{th}}$. The threshold must be chosen such that the end results do not depend on its particular value. 

In the second step, the Hamiltonian $H$ can be safely diagonalized in the orthonormal basis generated in the first step. Since ${\bold N}^{\tilde{\sigma}}$ is a Hermitian positive-definite matrix, the basis transformation can be written as
\begin{align}
{\mathbf N}^{\tilde{\sigma}} &=  {\mathbf S}^{\tilde{\sigma}\dagger} \breve{{\mathbf N}}^{\tilde{\sigma}} {\mathbf S}^{\tilde{\sigma}} \, ,\label{diago_Nkernel}
\end{align}
where ${\mathbf S}^{\tilde{\sigma}}$ is a unitary matrix and where $\breve{{\mathbf N}}^{\tilde{\sigma}}$ is diagonal with strictly positive eigenvalues. Defining 
\begin{align}
{\mathbf G}^{\tilde{\sigma}} &\equiv  
{\mathbf S}^{\tilde{\sigma}\dagger} 
\left(\breve{{\mathbf N}}^{\tilde{\sigma}}\right)^{-1/2} 
{\mathbf S}^{\tilde{\sigma}} 
\, , \label{change_basis}
\end{align}
and only keeping the rows of ${\mathbf S}^{\tilde{\sigma}}$ corresponding to eigenvalues of $\breve{{\mathbf N}}^{\tilde{\sigma}}$ larger than $\epsilon_{\text{th}}$, HWG's equation (Eq.~31 of Paper I) is transformed into the associated orthonormal basis and becomes
\begin{align}
\sum_{q} \breve{H}^{\tilde{\sigma}}_{p0q0} \, \breve{f}^{\tilde{\sigma}}_{\mu}(q) &=  {\cal E}^{\tilde{\sigma}}_{\mu} \, \breve{f}^{\tilde{\sigma}}_{\mu}(p) \, , \label{HWG_equation_transformed}
\end{align}
with
\begin{subequations}
\begin{align}
\breve{{\mathbf H}}^{\tilde{\sigma}} &\equiv  {\mathbf G}^{\tilde{\sigma}\dagger} {\mathbf H}^{\tilde{\sigma}} {\mathbf G}^{\tilde{\sigma}} \, , \label{change_basis_H} \\
{\mathbf f}^{\tilde{\sigma}} &\equiv  {\mathbf G}^{\tilde{\sigma}} \breve{{\mathbf f}}^{\tilde{\sigma}}  \, . \label{change_basis_f}
\end{align}
\end{subequations}
The solutions $\{\breve{f}^{\tilde{\sigma}}_{\mu}(q); q \in \text{set}\}$ play the role of orthonormal collective wave functions as a function of $q$ that can be interpreted as probability amplitudes. Left-multiplying Eq.~\ref{HWG_equation_transformed} by $\breve{f}^{\tilde{\sigma}\ast}_{\mu}(q)$, one can thus decompose the PGCM energy in terms of contributions associated with each deformation $q$
\begin{align}
{\cal E}^{\tilde{\sigma}}_{\mu} &= \sum_q  \breve{h}^{\tilde{\sigma}\ast}_{\mu}(q) \, \breve{f}^{\tilde{\sigma}}_{\mu}(q) \equiv \sum_q   e^{(0+1)}_{0}(q) \label{EPGCM_expanded}   \, ,
\end{align}
with
\begin{align}
\breve{h}^{\tilde{\sigma}}_{\mu}(q) &\equiv \sum_p  \breve{H}^{\tilde{\sigma}}_{q0p0} \, \breve{f}^{\tilde{\sigma}}_{\mu}(p) \label{EPGCM_expanded}   \, .
\end{align}
Note that, as done in Paper I, a similar decomposition of the PGCM energy can be achieved prior to diagonalizing the norm kernel.

\section{Memory optimization}

The storage of the interaction matrix elements necessary to perform \textit{ab initio} calculations in large computational bases is challenging. Several methods exist to reduce this memory burden. For example, one most commonly takes advantage of the rotational symmetry to store  matrix elements in \(J\)-coupled form. However, this storage is not well adapted to PGCM calculations based on symmetry breaking HFB states where the contraction of the interaction with rotated density matrices need to be performed in \(m\)-scheme. In the present Appendix, the workflow to calculate a Hamiltonian kernel while optimizing memory and runtime is detailed.

\subsection{$J$-coupling scheme}

In the present calculations, the one-body Hilbert space is spanned by spherical harmonic oscillator eigenstates that are labelled by 5 quantum numbers
\begin{equation}
    k \equiv (n_k, l_k, j_k, m_k, t_k) \, ,
\end{equation}
where \(n_k\) denotes the radial quantum number, \(l_k\) the orbital angular momentum, \(j_k\) the total angular momentum, \(m_k\) its projection and \(t_k\) the isospin projection. 

Introducing the reduced index
\begin{equation}
    \tilde k \equiv (n_k, l_k, j_k, t_k) \, ,
\end{equation}
and building the \(m\)-scheme, i.e. tensor-product, basis of the two-body Hilbert space according to
\begin{equation}
    \ket{k_1k_2} \equiv \ket{k_1} \otimes \ket{k_2} \, ,
\end{equation}
the \(J\)-coupled two-body basis is obtained through
\begin{equation}
    \ket{\tilde k_1\tilde k_2 JM} \equiv 
    \inv{1+\delta_{\tilde k_1\tilde k_2}} \sum_{m_1m_2} C^{JM}_{j_{k_1}m_{k_1}j_{k_2}m_{k_2}} \ket{k_1k_2} \, ,
\end{equation}
where Clebsch-Gordan coefficients have been introduced. Conversely, uncoupled basis states can be expanded on \(J\)-coupled ones via
\begin{equation}
    \ket{ k_1 k_2 } \equiv 
    ({1+\delta_{\tilde k_1\tilde k_2}}) \sum_{JM} C^{JM}_{j_{k_1}m_{k_1}j_{k_2}m_{k_2}} \ket{\tilde k_1 \tilde k_2 JM} \, .
\end{equation}

The two-body interaction being invariant under rotation, Wigner-Eckart theorem implies that its matrix elements are diagonal in \((J, M)\) when expressed in the \(J\)-coupled basis. Furthermore, matrix elements can be factorized in terms of a geometrical factor and a reduced tensor independent of \(M\). Therefore, only the reduced tensor is stored in memory. 

\subsection{Contractions with one-body density matrices}

When the PGCM solely involves spherically invariant HFB states, all necessary contractions of the two-body interaction with off-diagonal one-body density matrices associated with pairs of HFB vacua can be conveniently worked out in the \(J\)-coupled two-body basis in a way that only involves the reduced tensor. When employing deformed HFB states and projecting onto good angular momentum, interaction matrix elements must however be expressed in the uncoupled basis prior to performing the contractions. Two strategies are then possible
\begin{itemize}
    \item Uncouple \(J\)-coupled matrix elements prior to the calculation and work with the corresponding set of uncoupled matrix elements. This however induces a large memory requirement.
    \item Decouple interaction matrix elements on the fly, thus reducing the storage workload while substantially increasing the runtime.
\end{itemize}
In order to tackle this problem efficiently, a workflow that decouples each matrix element only once has been devised, thus reducing drastically both memory and runtime requirements in a way that is easily parallelized. The workflow is as follows
\begin{enumerate}
    \item Pre-compute all off-diagonal one-body density matrices.
    \item Split the initial one-body basis into subsets of states carrying the same quantum numbers \((m,\pi)\).
    \item Select $\left[(
    (m_1,\pi_1), 
    (m_2,\pi_2),
    (m_3,\pi_3),
    (m_4,\pi_4) 
    \right]$.
    \begin{enumerate}
        \item Decouple the sub-part of the interaction characterized by this combination of quantum numbers and store it contiguously in memory.
        \item Perform the contraction of the interaction sub-part with the corresponding sub-blocks of the off-diagonal one-body density matrices. This part can be completely vectorized since the decoupled interaction is stored contiguously in memory.
    \end{enumerate}
    \item Go back to 3. until all combinations of quantum numbers have been exhausted.
    \end{enumerate}
The loop in step 3 can be easily parallelized. Except for the overhead associated with the storage of all off-diagonal one-body density matrices, the memory consumption scales linearly with the number of cores.

\section{Evaluation of the norm overlap}

As discussed in Paper I, the overlap between a left Bogoliubov vacuum and a rotated right Bogoliubov vacuum can be evaluated according to~\cite{bertsch12} 
\begin{strip}
\begin{align}
  \braket{\Phi(p)}{\Phi(q;\theta)} 
  &=
  (-1)^n
  \frac{\det(C^*(p))\det(C(q))}{\prod_k^n v_{k}(p)v_{k}(q)} \mathrm{pf} \left[
    \begin{pmatrix}
      V(p)^T U(p)     &   V^T(p)  {\bold r}^T(\theta) V^*(q)\\
      -V(q)^\dag  {\bold r}(\theta) V(p)  &   U^\dag(q) V^*(q)
    \end{pmatrix}
    \right] \, ,
    \label{eq:overlap}
\end{align}
where the pfaffian of a symplectic matrix and Bloch-Messiah-Zumino's decompositions~\cite{RiSc80} of the Bogoliubov transformations of the left and (unrotated) right states have been invoked. In Eq.~\eqref{eq:overlap}, $2n$ denotes the dimension of ${\cal H}_1$. This expression is however not numerically stable as it amounts to taking the ratio of two vanishing quantities. A way to circumvent this difficulty has been proposed in Ref.~\cite{Carlsson_2021}, but a simpler alternative consists of rescaling the matrix before evaluating the pfaffian. For any \(2n\times 2n\) skew-symmetric matrix \(A\) and real scalar \(\lambda\), one has
\begin{equation}
    \mathrm {pf}(\lambda A) = \lambda ^ n\mathrm {pf}( A).
\end{equation}
Therefore, Eq. \eqref{eq:overlap} can be rewritten as
\begin{align}
  \braket{\Phi(p)}{\Phi(q;\theta)} 
  &=
  (-1)^n
  \det(C^*(p))\det(C(q))  \mathrm{pf} \left[\inv{\sqrt[n]{\prod_k^n v_{k}(p)v_{k}(q)}}
    \begin{pmatrix}
      V(p)^T U(p)     &   V^T(p)  {\bold r}^T(\theta) V^*(q)\\
      -V(q)^\dag  {\bold r}(\theta) V(p)  &   U^\dag(q) V^*(q)
    \end{pmatrix}
    \right] \, , \label{eq:overlap_alt}
\end{align}
\end{strip}
which is well-behaved numerically. The numerical library Pfapack \cite{Wimmer_2012} is used to compute the pfaffian.

\section{Charge density distribution}
\label{ch_density}

Generically speaking, the electromagnetic charge density operator is expressed as an expansion in many-body operators acting on nucleonic degrees of freedom. 
These operators not only account for the point distribution of protons but also for their own charge distribution, along with the one of neutrons, and for charge distributions associated with the light charged mesons they exchange.

In practice, the charge density distribution\footnote{An additional relativistic correction that depends on spin-orbit terms, $\rho_{{\rm ch}}^{{\rm ls}}$, is sometimes considered. Given that proton and neutron spin-orbit contributions largely cancel out in $N=Z$ nuclei, this term is omitted in the present calculation of $^{20}$Ne.} is usually computed as~\cite{Bertozzi72,Chandra76,Brown79}
\begin{equation}
\rho_{{\rm ch}}(r) = \rho_{{\rm ch}}^{{\rm p}}(r) + \rho_{{\rm ch}}^{{\rm n}}(r),
\label{charge_density}
\end{equation}
where $\rho_{{\rm ch}}^{{\rm p}}$ ($\rho_{{\rm ch}}^{{\rm n}}$) is determined by folding the point-proton (point-neutron) density with the finite charge distribution of the proton (neutron).
Following Ref.~\cite{Chandra76}, the latter are included by parameterizing proton and neutron charge form factors as a linear superposition of Gaussians 
\begin{equation}
G_{{\rm p}/{\rm n}}(r) = \sum_i \frac{\theta^{{\rm p}/{\rm n}}_i}{[\pi (r^{{\rm p}/{\rm n}}_i)^2]^{3/2}} \, \text{e}^{-(r/r_i^{{\rm p}/{\rm n}})^{2}} \; ,
\label{pnFF}
\end{equation}
whose widths $r^{{\rm p}/{\rm n}}_i$ and relative weights $\theta^{{\rm p}/{\rm n}}_i$ are adjusted to reproduce electron scattering data. Three and two Gaussians are sufficient to reproduce proton and neutron form factors, respectively, with parameters\footnote{The proton r.m.s. radius resulting from this parameterization is $\langle R^2_{\rm p} \rangle^{1/2} = 0.88 ~\text{fm}$.
This is consistent with the values reported in the older CODATA evaluations (e.g. the 2010 evaluation~\cite{CODATA2010}, $\langle R^2_{{\rm p}} \rangle^{1/2} = 0.8775(51)\,\text{fm}$), but overestimates the value found in more recent evaluations ($\langle R^2_{{\rm p}} \rangle^{1/2} = 0.8414(19)\,\text{fm}$, adopted from the 2014 evaluation~\cite{CODATA2014} on).
A smaller value of the proton r.m.s. charge radius would lead to less smoothing of the point-proton distribution. Given the small difference between the possible values of $\langle R^2_{\rm p} \rangle^{1/2}$, however, this would be hardly noticeable in the final charge density curves.} given in Tab.~\ref{chandra}.
\begin{table}
\centering
\renewcommand{\arraystretch}{1.4}
\begin{tabular}{|c|c|c|}
\cline{1-3}
 & proton & neutron  \\
\cline{1-3}
$\theta^{{\rm p}/{\rm n}}_1$ & 0.506 & 1  \\
$\theta^{{\rm p}/{\rm n}}_2$ & 0.328 & -1  \\
$\theta^{{\rm p}/{\rm n}}_3$ & 0.166 & - \\
$(r_1^{{\rm p}/{\rm n}})^2 \, [\text{fm}^2]$ & 0.432 & 0.469 \\
$(r_2^{{\rm p}/{\rm n}})^2 \, [\text{fm}^2]$ & 0.139 & 0.546 \\
$(r_3^{{\rm p}/{\rm n}})^2 \, [\text{fm}^2]$ & 1.526 & - \\
\cline{1-3}
$\langle R_{{\rm p}/{\rm n}}^{2} \rangle \, [\text{fm}^2]$ & 0.775 & -0.116 \\
\cline{1-3}
\end{tabular}
\caption{Proton and neutron parameters entering the Gaussian expansion~\eqref{pnFF}. Taken from Ref.~\cite{Chandra76}. The resulting mean-square charge radii are also reported.}
\label{chandra}
\renewcommand{\arraystretch}{1}
\end{table}
Convoluting Eq.~\eqref{pnFF} with point-proton and point-neutron distributions $\rho_{{\rm p}}$ and $\rho_{{\rm n}}$ yields the two contributions to the nuclear charge density~\cite{Brown79}
\begin{subequations}
\begin{equation}
\rho_{{\rm ch}}^{{\rm p}}(r) = \hspace{-.05cm}  \sum_{i=1}^{3} \frac{\theta_i^{{\rm p}}}{r_i^{{\rm p}}\sqrt{\pi}} \hspace{-.1cm} \int\limits_{0}^{+\infty} \hspace{-.1cm}  dr' \frac{r'}{r} \rho_{{\rm p}}(r')\left\lbrack \text{e}^{-\left(\frac{r-r'}{r_i^{{\rm p}}}\right)^{2}} \hspace{-.15cm}  - \text{e}^{-\left(\frac{r+r'}{r_i^{{\rm p}}}\right)^{2}} \right\rbrack ,
\end{equation}
\begin{equation}
\rho_{{\rm ch}}^{{\rm n}}(r) = \hspace{-.05cm}  \sum_{i=1}^{2} \frac{\theta_i^{{\rm n}}}{r_i^{{\rm n}}\sqrt{\pi}} \hspace{-.1cm} \int\limits_{0}^{+\infty} \hspace{-.1cm}  dr' \frac{r'}{r} \rho_{{\rm n}}(r')\left\lbrack \text{e}^{-\left(\frac{r-r'}{r_i^{{\rm n}}}\right)^{2}} \hspace{-.15cm}  - \text{e}^{-\left(\frac{r+r'}{r_i^{{\rm n}}}\right)^{2}} \right\rbrack .
\end{equation}
\label{pnchden}
\end{subequations}

Finally, one needs to correct for spurious center-of-mass contamination and include the Darwin-Foldy relativistic correction. Assuming that the center-of-mass wave function factorizes in the ground-state of a harmonic oscillator Hamiltonian characterized by the frequency $\tilde{\omega}$, the inclusion of these two corrections can be performed at the price of proceeding to the replacement~\cite{Negele70,Chandra76}
\begin{equation} 
r^2_i \longrightarrow r^2_i  - \frac{b^2}{A} + \frac{1}{2}\left(\frac{\hbar}{m}\right)^2
\end{equation}
in Eqs.~\eqref{pnchden},  where $m$ is the nucleon mass, hence $\hbar/m= 0.21$\,fm, and $b^2=(m \, \hbar \, \tilde{\omega})^{-1}$.
Employing Bethe's formula~\cite{Negele70}, the latter term can be approximated with $b^2 \approx A^{1/3} \,\,\text{fm}^2$. Let us note that, for $^{16}$O, such an approximation is consistent with the value of $\hbar\tilde{\omega}$ found in Ref.~\cite{Hagen09} and is thus safe to use in present calculations of $^{20}$Ne.

\end{appendices}
\interlinepenalty=10000
\bibliography{bibliography.bib}

\end{document}